\documentclass[%
reprint,
twocolumn,
superscriptaddress,
amsmath,
amssymb,
aps,
prb,
]{revtex4-1} % chktex 8
\usepackage[T1]{fontenc}    % represent special characters like ö correctly
\usepackage[utf8]{inputenc} % use utf8
\usepackage[english]{babel} % use English

\usepackage{amsmath}
\usepackage{amssymb}
\usepackage{mathrsfs}   % \mathcal
\usepackage{mathtools}  % extends amsmath
\usepackage{dsfont}     % for double stroked number/letters

\usepackage{graphicx,xcolor}
\graphicspath{{fig/}}

\usepackage{microtype}  % fix margin (lines ending with `.`)

\usepackage{siunitx}    % units
\usepackage{suffix}     % define stared commands
\usepackage{physics}    % many useful commands for physics, read the manual

\usepackage{scalerel}
\usepackage{tikz}
\usetikzlibrary{positioning}
\usetikzlibrary{svg.path}

\definecolor{orcidlogocol}{HTML}{A6CE39}
\tikzset{%
  orcidlogo/.pic={%
    \fill[orcidlogocol] svg{M256,128c0,70.7-57.3,128-128,128C57.3,256,0,198.7,0,128C0,57.3,57.3,0,128,0C198.7,0,256,57.3,256,128z};
    \fill[white] svg{M86.3,186.2H70.9V79.1h15.4v48.4V186.2z}
                 svg{M108.9,79.1h41.6c39.6,0,57,28.3,57,53.6c0,27.5-21.5,53.6-56.8,53.6h-41.8V79.1z M124.3,172.4h24.5c34.9,0,42.9-26.5,42.9-39.7c0-21.5-13.7-39.7-43.7-39.7h-23.7V172.4z}
                 svg{M88.7,56.8c0,5.5-4.5,10.1-10.1,10.1c-5.6,0-10.1-4.6-10.1-10.1c0-5.6,4.5-10.1,10.1-10.1C84.2,46.7,88.7,51.3,88.7,56.8z};
  }
}

\newcommand\orcidicon[1]{\href{https://orcid.org/#1}{\mbox{\scalerel*{
\begin{tikzpicture}[yscale=-1,transform shape]
\pic{orcidlogo};
\end{tikzpicture}
}{M}}}}

\definecolor{linkcolor}{RGB}{0, 68, 136}
\usepackage[unicode,colorlinks,allcolors=blue]{hyperref}
\usepackage[all]{hypcap}
\usepackage[capitalize]{cleveref}   % use \cref for equation, figures, define commands as needed

\newcommand{\I}{\mathds{1}}
\newcommand{\mat}[1]{\boldsymbol{#1}}
\newcommand{\transp}{\intercal}
\renewcommand{\oc}{\hat{c}^{\vphantom{\dagger}}}
\newcommand{\ocd}{\hat{c}^\dagger}
\DeclareMathOperator{\config}{conf}
\newcommand{\ConfSet}{{\mathbb{C}}}
\newcommand{\RestConfSet}{{\mathbb{C}\vert_{\{c^{\alpha}\}}}}
\newcommand{\SiteSet}{\mathbb{S}}
\DeclareMathOperator{\Expval}{\mathbb{E}}
\newcommand{\cmp}[1]{\mat{\underline{\mathrm{#1}}}}
\WithSuffix\newcommand\cmp*[1]{\underline{\mathrm{#1}}}
\newcommand{\scmp}[1]{\cmp{\tilde{#1}}} % scaled version
\WithSuffix\newcommand\scmp*[1]{\cmp*{\tilde{#1}}} % scaled version
\newcommand{\indi}[1]{1_{#1}} % indicator function
\newcommand{\cmpind}[1]{\indi{\SiteSet^{#1}}}
\newcommand{\nodesize}{0.6cm}
\newcommand{\olegsize}{0.3cm}

\DeclareMathOperator{\e}{e}
\DeclareMathOperator{\Fourier}{F}
\newcommand{\Gtret}{{G}^{\text{ret}}}
\newcommand{\tmat}{\mat{\mathsf{t}}}
\newcommand{\Gloc}{\cmp{g}_{\text{loc}}}
\WithSuffix\newcommand\Gloc*{\cmp*{g}_{\text{loc}}}
\newcommand{\CmpA}{\mathsf{A}}
\newcommand{\CmpB}{\mathsf{B}}
\newcommand{\cA}{c^{\CmpA}}
\newcommand{\cB}{c^{\CmpB}}
\newcommand{\vA}{\cmp*{v}^\CmpA}
\newcommand{\vB}{\cmp*{v}^\CmpB}
\newcommand{\UA}{\cmp*{U}^\CmpA}
\newcommand{\UB}{\cmp*{U}^\CmpB}
\newcommand{\cmpT}{\cmp{T}}
\WithSuffix\newcommand\cmpT*{\cmp*{T}}
\newcommand{\tAA}{\cmpT*^{\CmpA\CmpA}}
\newcommand{\tBB}{\cmpT*^{\CmpB\CmpB}}
\newcommand{\tAB}{\cmpT*^{\CmpA\CmpB}}
\DeclareMathOperator{\sign}{sign}

\begin{document}
\title{Dynamical mean-field theory of the Anderson--Hubbard model with local and non-local disorder in tensor formulation}
\author{A. Weh \orcidicon{0000-0003-1213-2919}} % chktex 8
\email{andreas.weh@physik.uni-augsburg.de}
\affiliation{Theoretical Physics II,  Institute of Physics, University of Augsburg, 86135 Augsburg, Germany}
\author{Y. Zhang \orcidicon{0000-0003-3041-8371}} % chktex 8
\affiliation{Kavli Institute of Theoretical Sciences, University of Chinese Academy of Sciences, Beijing, 100190, China}
\author{A. Östlin \orcidicon{0000-0003-1225-1635}} % chktex 8
\affiliation{Theoretical Physics III, Center for Electronic Correlations and Magnetism, Institute of Physics, University of Augsburg, 86135 Augsburg, Germany}
\affiliation{Augsburg Center for Innovative Technologies, University of Augsburg, 86135 Augsburg, Germany}
\author{H. Terletska}
\affiliation{Department of Physics and Astronomy, Middle Tennessee State University, Murfreesboro, TN 37132, USA}
\author{D. Bauernfeind}%
\affiliation{Center for Computational Quantum Physics, Flatiron Institute, 162 5th Avenue, New York, NY 10010, USA}
\author{K.-M.~Tam}
\affiliation{Department of Physics \& Astronomy, Louisiana State University, Baton Rouge, LA 70803, USA}
\affiliation{Center for Computation \& Technology, Louisiana State University, Baton Rouge, LA 70803, USA}
\author{H. G. Evertz \orcidicon{0000-0002-2037-0725}} % chktex 8
\affiliation{Institute of Theoretical and Computational Physics, Graz University of Technology, 8010 Graz, Austria}
\author{K. Byczuk \orcidicon{0000-0003-2409-3427}} % chktex 8
\affiliation{Institute of Theoretical Physics, Faculty of Physics, University of Warsaw, ul. Pasteura 5, 02-093 Warszawa, Poland}
\author{D. Vollhardt}
\affiliation{Theoretical Physics III, Center for Electronic Correlations and Magnetism, Institute of Physics, University of Augsburg, 86135 Augsburg, Germany}
\author{L. Chioncel \orcidicon{0000-0003-1424-8026}} % chktex 8
\affiliation{Theoretical Physics III, Center for Electronic Correlations and Magnetism, Institute of Physics, University of Augsburg, 86135 Augsburg, Germany}
\affiliation{Augsburg Center for Innovative Technologies, University of Augsburg, 86135 Augsburg, Germany}
\date{\today}

\begin{abstract}
To explore correlated electrons in the presence of local and non-local disorder, the Blackman--Esterling--Berk method for averaging over off-diagonal disorder is implemented into dynamical mean-field theory using tensor notation.
The impurity model combining disorder and correlations is solved using the recently developed fork tensor-product state solver, which
allows one to calculate the single particle spectral functions on the real-frequency axis.
In the absence of off-diagonal hopping,
we establish exact bounds of the spectral function of the non-interacting Bethe lattice with coordination number $Z$.
In the presence of interaction, the Mott insulating paramagnetic phase of the one-band Hubbard model is computed at zero temperature in alloys with site- and off-diagonal disorder.
When the Hubbard \(U\) parameter is increased transitions from an alloy band-insulator through a correlated metal into a Mott insulating phase are found to take place.
\end{abstract}

\maketitle

%***************************************************
\section{Introduction}
%*************************************************

The electronic structure and transport properties of real materials are strongly influenced by the Coulomb interaction between the electrons and the presence of disorder~\cite{Lee+Rama_1985,Mott_1990,Belitz_1994}. In particular, both electronic correlations and randomness are driving forces behind a transition from a metallic to an insulating state due to the localization of electrons (``metal-insulator transition'' (MIT)).
While the Mott--Hubbard MIT is caused by the repulsive interaction between the electrons~\cite{Mott_1949,Mott_1990,Imada_1998}, the Anderson MIT is a result of coherent backscattering of non-interacting electrons from randomly distributed impurities~\cite{Anderson_1958,Lee+Rama_1985,Evers+Mirlin_2008,Book-50years_2010}. The interplay between interactions and static
disorder gives rise to many unusual and often unexpected phenomena~\cite{Lee+Rama_1985,Belitz_1994,Abrahams_2001,Book-50years_2010,Book-Dobrosl_2013,Huse_2015,Altman_2019}.
The simplest model of disordered interacting electrons is the Anderson--Hubbard model, obtained by supplementing a single-band Hubbard model with local and/or non-local disorder.
If the disorder acts only locally, i.e., via random local potentials (``diagonal disorder''), this model is able to describe substitutionally disordered binary alloys.
However, in general disorder also affects the amplitudes for hopping between two sites --- especially when the bandwidths of the host and dopant are very different --- leading to additional ``off-diagonal disorder''.
In analytic calculations local disorder is easier to treat and was studied extensively~\cite{Ziman_1979}.
In particular, the coherent potential approximation (CPA)~\cite{p_soven_67,tayl.67,Velicky_1968,Elliott_1974,econ.06,Gonis_1992}
provides the best single-site approximation for non-interacting systems with local disorder.
For that reason the simultaneous investigation of diagonal disorder within the CPA and of interacting electrons with local (Hubbard) interaction within the dynamical mean-field theory (DMFT)~\cite{me.vo.89,Georges+Kotliar_1992,ge.ko.96,ko.vo.04}
fit together particularly well, since both DMFT~\cite{me.vo.89,ge.ko.96} and CPA~\cite{Vlaming_1992,ja.vo.92} become exact in the limit of infinite spatial dimensions or lattice coordination number.

The treatment of the Anderson--Hubbard model with off-diagonal disorder received somewhat less attention~\cite{Bhatt_1989,Dob_1993,Dob_1994,Ulmke_1997,Dent_2001,Miranda_2013}.
In particular, Dobrosavljevi\'{c} and Kotliar~\cite{Dob_1993,Dob_1994} investigated this model within DMFT by employing a functional integral representation for quantum averages and the replica method for disorder averaging.
Thereby they were able to study Hubbard models with arbitrary disorder on Bethe lattices, as well as models on an arbitrary lattice with a special distribution of the off-diagonal disorder.
In this way they studied the formation of local moments and the Mott transition in disordered systems.
In 1971 Blackman, Esterling, and Berk (BEB)~\cite{Blackman_1971,Esterling_1975} showed that off-diagonal disorder can, in principle, be incorporated into the CPA framework, such that both diagonal and off-diagonal disorder are tractable within a single–site approximation.
In the absence of electronic interactions the BEB formalism was incorporated into the dynamical cluster approximation within the typical medium cluster theory~\cite{te.ek.14}
and applied to multi-band systems~\cite{Jarrell_2016}.

In our paper we extend these investigations of disordered systems by including a local (``Hubbard'') interaction between the electrons.
To this end we investigate the Anderson--Hubbard model with diagonal and off-diagonal disorder within the CPA in the BEB formulation~\cite{Blackman_1971,Esterling_1975} in a tensor formulation~\cite{ko.ve.97,ko.ve.98}, while correlation effects are treated within the DMFT~\cite{ge.ko.96}.
We compute the spectral function and discuss the occurrence of the MIT in the presence of diagonal and off-diagonal disorder.
For these purposes an accurate zero-temperature many-body solver on the real energies is used,
the so-called fork tensor-product state solver~\cite{ba.zi.17}.
We observe successive alloy-insulator to metal and metal to Mott-insulator transitions with increasing values of the Hubbard \(U\) parameter.
A similar transition scenario was previously discussed for models with diagonal disorder solved within DMFT, but using finite temperature solvers such as the Hirsch-Fye algorithm~\cite{Ulmke_1995} or the perturbative non-crossing approximation~\cite{lo.ha.06}.
Contrary to the diagonal disorder model in which the CPA solution provides a common bandwidth for all alloy components, the presence of off-diagonal elements causes the formation of different effective bandwidths for alloy components.
The paper is organized as follows.
In \cref{sec:G-BEB} the BEB theory of the multi-component Anderson--Hubbard model is formulated,
and the computational scheme is discussed.
In \cref{sec:results} numerical results for a single-band Bethe lattice are presented.
Comparison with earlier results on DMFT+CPA allow us to identify effects specifically due to off-diagonal disorder.
Finally, conclusions and a summarizing discussion are presented in \cref{sec:summary}.

\section{Blackman--Esterling--Berk theory for the Anderson--Hubbard model}\label{sec:G-BEB}
In the simplest case an alloy consists of two types of atoms, \(\CmpA\) and \(\CmpB\), with diagonal substitutional disorder,
such that only the site-diagonal elements of the Hamiltonian vary stochastically according to the atomic species occupying the given site~\cite{p_soven_67,tayl.67,Velicky_1968}.
The physical quantities of interest, for instance the spectral function, are those averaged over the possible disorder realizations.
Therefore, the idea of the CPA is to replace the ensemble with random configurations by a periodic system with ``average'' atoms,
whose properties are determined self-consistently.
The CPA finds a natural description in the language of scattering theory.
Assuming the origin to be occupied by atoms of type \(\CmpA\) or \(\CmpB\) and all other sites by average atoms,
the scattering by the atom at the origin is easily computed.
The self-consistency condition of the CPA,
which expresses that the scattering at the origin vanishes on average,
then allows one to compute the coherent Green's function for the average atom.
The BEB formalism~\cite{Blackman_1971,Esterling_1975} is a generalization of the CPA such that it becomes applicable also to off-diagonal disorder.
Similar to CPA it was formulated for a tight-binding model in which the hopping matrix elements depend on the species of atoms occupying the two sites connected by the hopping.
For the above binary alloy example,
the BEB hopping matrix elements are \(t_{\CmpA\CmpA}, t_{\CmpA\CmpB}\) and \(t_{\CmpB\CmpB}\).
When the hopping matrix elements are equal the BEB formalism reduces to the CPA~\cite{p_soven_67,tayl.67},
and for the binary alloy case the scalar CPA equation becomes a $2\times2$ matrix equation.
An in-depth analysis of the BEB method was performed in a tight-binding formalism by Gonis and Garland~\cite{go.ga.77} using locators, propagators, and a variational technique proving the analyticity of the BEB-CPA Green's function.
A realistic multiband formulation of the BEB-CPA was introduced more than three decades ago by Papaconstantopoulos, Gonis, and Laufer~\cite{pa.go.89}.
Then Koepernik \emph{et al.}~\cite{ko.ve.97} developed the BEB-CPA extension within a full potential local-orbital approach, and more recently a similar implementation was made within a pseudopotential approach~\cite{he.he.17}.
Shvaika~\cite{shva.03} found a connection between the Falicov--Kimball model with correlated hopping and the BEB-CPA by rewriting the Hamiltonian as a $2\times2$ matrix.
For Hamiltonians with interactions the BEB-CPA was employed by Burdin and Fulde~\cite{Burdin_2007} to study the interplay between the Kondo effect and disorder.
In an attempt to address localization in strongly disordered electronic systems,
the typical medium theory~\cite{Dob_1993} was combined with the dynamical cluster approximation including effects induced by off-diagonal disorder~\cite{te.ek.14}.
However, this approach did not include electronic interactions.
In the present paper we extend the BEB formalism to \emph{interacting} electrons such that it can be applied to
the multi-component Anderson--Hubbard model;
localization effects will not be addressed.
The model is defined in \cref{sec:model}.
The corresponding DMFT equations are solved using the recently developed fork tensor-product state solver~\cite{ba.zi.17}.

\subsection{Configurational averages and notation}
In the conventional approach to systems with random variables (diagonal and/or off-diagonal) the Green's function is first expanded and then an average over an appropriate set of terms is performed.
By contrast, the BEB method treats both diagonal- and off-diagonal randomness on equal footing by employing an extended representation, which will be discussed below.
The Green's functions are then evaluated using conventional expansion techniques.
The formalism introduced by Koepernik \emph{et al.}~\cite{ko.ve.97,ko.ve.98} is particularly suitable for the BEB approach.
For this reason, we adopt the notation introduced in Refs.~\onlinecite{ko.ve.97,ko.ve.98}.
We consider an alloy consisting of \(M\) types of atoms (``alloy components'') denoted by the index \(\alpha\).
Every lattice site \(i\) is uniquely mapped to a particular component \(\alpha\) as expressed by
\begin{equation}\label{eq:mapping}
  i \mapsto \alpha.
\end{equation}
While this notation corresponds, in principle, to that of Ref.~\onlinecite{ko.ve.98},
we reverse the direction of the arrow to focus on the alloy components rather than the lattice sites.
To address multiple sites we use the notation
\((i,j \mapsto \alpha, \beta) \coloneqq (i \mapsto \alpha) \land (j \mapsto \beta)\).
We refer to a specific mapping of the \(N\) lattice sites to the \(M\) components as a ``configuration'' (\(\config\)) or ``disorder realization'',
and denote the set of all possible configurations by \({\ConfSet} = \{\config\}\).
As the specific configuration of a sample measured in an experiment is unknown,
we average over all possible configurations.
Since the concentrations \(c^{\alpha}\) of the different components \(\alpha\) are assumed to be known
we restrict the average to configurations with these concentrations,
denoted by \({\ConfSet}\vert_{\{c^{\alpha}\}}\).
In the absence of additional information we further assume that the probability of all physical configurations is the same:
\begin{equation}
  P(\config) = 1/\abs{\RestConfSet}
  \quad \forall \config \in \RestConfSet.
\end{equation}
For a random variable \(X\), the stochastic average over all physical configurations is the weighted sum
\begin{equation}\label{expectation}
  \Expval(X)
  = \mkern-12mu \sum_{\config \in \RestConfSet}\mkern-18mu P(\config) x_{\config}
  = \frac{1}{\abs{\RestConfSet}} \sum_{\config \in \RestConfSet}\mkern-18mu  x_{\config}.
\end{equation}
This situation corresponds to the case of substitutional disorder.
\begin{table*}[hbt]
  \begin{ruledtabular}
    \begin{tabular}{rlrl}
      \(i,j, \dots\)   & site indices
      & \(c^{\alpha}\) & concentration of component \(\alpha\)
      \\%%
      \(\alpha, \beta, \dots\)  & alloy components
      & $\cmp*{v}^{\alpha} $ &   on-site energy of component $\alpha$
      \\%%
      \(i \mapsto \alpha\) & mapping of a site to an alloy component
      & $\cmp*{U}^{\alpha} $ &   Hubbard parameter of component $\alpha$
      \\%%
      \(\SiteSet^{\alpha}\) & set of sites occupied by component \(\alpha\)
      &$\cmp*{t}^{\alpha \beta}(\abs{\vb{r}_{i} - \vb{r}_{j}})$ &
       hopping between component \(\alpha\) and \(\beta\)
      \\%%
      \(\RestConfSet\) & set of configurations (disorder realization)
      &$\cmp*{H}_{ij}^{\alpha \beta}$ & extended Anderson--Hubbard Hamiltonian matrix
      \\%%
      \(X\) & scalar random variable
      &\({\cmp*{\eta}}^{\alpha}_{ij}\) & indicator tensor, represented as equivalent matrix
      \\%%
      \(\Expval(X)\) & Expectation value defined according to \cref{expectation}
      &\(\cmp{\chi}\) & projector onto specific disorder configuration
      \\%%
      &&\(\cmpT*^{\alpha \beta}\) & dimensionless hopping parameter\\
      &&&describing hopping between components \(\alpha\) and \(\beta\)
    \end{tabular}
  \end{ruledtabular}
  \caption{Notations specific to the extended basis and Hamiltonian parameters used in BEB method.}\label{tab:disorder-symbols}
\end{table*}

\subsection{Multi-component Anderson--Hubbard model}\label{sec:model}
For a specific configuration the Anderson--Hubbard Hamiltonian reads
\begin{equation}
  \hat{H}
  =-\sum_{ij\sigma}t_{ij} \ocd_{i\sigma} \oc_{j\sigma}
  +\sum_{i\sigma}(v_i - \mu) \hat{n}_{i\sigma}
  + \sum_i U_i \hat{n}_{i \uparrow} \hat{n}_{i \downarrow},
\end{equation}
with the on-site energy \(v_i\), the local Hubbard interaction \(U_{i}\), and the amplitude \(t_{ij}\) for hopping between sites $i$ and $j$.
The hopping parameters are Hermitian \(t_{ij} = t^*_{ji}\) and off-diagonal, with \(t_{ii} =0\).
The Hamiltonian can be written in the compact matrix form
\begin{equation}\label{eq:hamiltonian_mat-form}
  \hat{H}
  = \sum_{\sigma} \mat{\oc}^{\dagger}_{\sigma} \mat{H}_{\sigma} \mat{\oc}_{\sigma}
  + \mat{\hat{n}}^{\transp}_{\uparrow} \mat{U} \mat{\hat{n}}_{\downarrow},
\end{equation}
where we introduced \(N \times 1\) matrices to represent the operators.
The rows of the matrix \(\mat{\oc_{\sigma}}\) are the annihilation operators \(\oc_{i \sigma}\),
and the rows of \(\mat{\hat{n}}_{\sigma}\) are the number operators \(\hat{n}_{i \sigma}\).
The one-particle Hamiltonian matrix reads \({(\mat{H})}_{ij} = - t_{ij} + \delta_{ij} (v_{i} - \mu)\).
Here and in the following we suppress the spin-index \(\sigma\) unless explicitly needed, to simplify the notation.
The local interaction is written as a matrix \({(\mat{U})}_{ij} = \delta_{ij} U_{i}\).
The magnitude of the hopping parameters \(t_{ij}\) depends on the alloy components located on sites \(i\) and \(j\), respectively,
which are referred to as ``terminal points''.
In the following we employ the ``terminal-point approximation''~\cite{ko.ve.97,ko.ve.98}
which assumes that parameters with terminal points \(i,j,k, \dots\) depend only on the components located at \(i,j,k, \dots\)
and not on the components surrounding these sites.
Thus, for a specific configuration (disorder realization)
every parameter \(v_{i}, U_{i}, t_{ij}\) takes a value depending on the component occupying the respective site or sites.
In the representation of the BEB we denote these configuration-specific values by an underline and a superscript indicating the component.
For instance, if site \(i\) is occupied by component \(\alpha\) (\(i \mapsto \alpha\))
the parameter \(v_{i}\) takes the value \(\cmp*{v}^{\alpha}\).
For \(i, j \mapsto \alpha, \beta\),
we have \(v_{i} = \cmp*{v}^\alpha\), \(U_{i} = \cmp*{U}^{\alpha}\), and \(t_{ij} = \cmp*{t}^{\alpha \beta}(\abs{\vb{r}_{i} - \vb{r}_{j}})\).
This will now be formalized.
Denoting the set of sites \(i\) occupied by the component \(\alpha\) by
\begin{equation}
  \SiteSet^{\alpha}
  \coloneqq \{i\vert i \mapsto \alpha\},
\end{equation}
the terminal point approximation can be expressed conveniently using the indicator function
\begin{equation}
  \indi{\SiteSet^{\alpha}}(i)
  \coloneqq \begin{cases}
    1 &\qif i \in \SiteSet^{\alpha},
    \\%%
    0 &\qif i \not\in \SiteSet^{\alpha}.
  \end{cases}
\end{equation}
The identity \(\sum_{\alpha}\cmpind{\alpha}(i)=1\) holds since
every site must be occupied by exactly one component.
Thus, the parameters read
\begin{equation}\label{eq:cmp-parameters}
\begin{aligned}
  v_{i}
  &= \sum_{\alpha} \cmpind{\alpha}(i)\cmp*{v}^{\alpha},
  \\%%
  t_{ij}
  &= \sum_{\alpha \beta} \cmpind{\alpha} (i)\cmp*{t}^{\alpha \beta}(\abs{\vb{r}_{i} - \vb{r}_{j}}) \cmpind{\beta}(j),
  \\%%
  H_{ij}
  &= \sum_{\alpha \beta} \cmpind{\alpha} (i)\cmp*{H}^{\alpha \beta}_{ij} \cmpind{\beta}(j),
  \\%%
  U_{i}
  &= \sum_{\alpha} \cmpind{\alpha}(i)\cmp*{U}^{\alpha},
\end{aligned}
\end{equation}
with \(\cmp*{H}_{ij}^{\alpha \beta} = \delta_{ij}\delta^{\alpha \beta} (\cmp*{v}^{\alpha} - \mu) - \cmp*{t}^{\alpha \beta}(\abs{\vb{r}_{i}- \vb{r}_{j}})\).
We further note,
that the conditional expectation value of the parameters equals the underlined component variables:
\begin{equation}
  \Expval(v_{i}\vert i \mapsto \alpha)
  = \cmp*{v}^{\alpha},\quad
  \Expval(t_{ij}\vert i,j \mapsto \alpha, \beta)
  = \cmp*{t}^{\alpha \beta}, \dots
\end{equation}
The dependence on the components is therefore shifted from the parameters into the indicator function \(\cmpind{\alpha}(i)\).
Depending on the component occupying a site, the indicator function selects the corresponding parameter from a finite set of choices.
We note that for elements diagonal in lattice sites \(i\),
one has \(\cmpind{\alpha}(i)\cmpind{\beta}(i) = \cmpind{\alpha}(i) \delta^{\alpha \beta}\), i.e.,
they are diagonal in the components.
In the following we refer to a quantity with multiple indices, which include both site and component indices, as a ``tensor''.

We introduce the indicator tensor
\begin{equation}\label{eq:indicator-tensor}
  {\cmp*{\eta}}_{ij}^{\alpha}
  = \cmpind{\alpha}(i) \delta_{ij}
  = \begin{tikzpicture}[baseline=(eta.base), minimum size=\nodesize, node distance=0.5cm]
    \node[rectangle, draw] (eta) at (0, 0) {\(\cmp{\eta}\)};
    \draw ([yshift=1ex]eta.west) -- ++ (-\olegsize, 0) node[left, minimum size=0cm] {\scriptsize \(\alpha\)};
    \draw ([yshift=-1ex]eta.west) -- ++ (-\olegsize, 0) node[left, minimum size=0cm] {\scriptsize \(i\)};
    \draw ([yshift=-1ex]eta.east) -- ++ (\olegsize, 0) node[right, minimum size=0cm] {\scriptsize \(j\)};
  \end{tikzpicture}.
\end{equation}
Graphically we represent this tensor as a box with legs as seen on the right-hand side of \cref{eq:indicator-tensor}.
The order of the tensor is given by the number of its legs, here three.
The upper leg carries the alloy component indices \(\alpha\),
and the lower legs correspond to the site indices \(i,j\).
Within our matrix notation this tensor is equivalent to a \(MN \times N\) matrix.
We group the left indices for sites \(i\) and components \(\alpha\),
or in the graphical notation the legs above each other.
In the following we refer to the \(MN\)-dimensional vector space of grouped sites and components as ``extended space''.
Matrix products in the extended space sum over the grouped \(MN\) elements for component and site indices;
they are equivalent to the tensor contraction of two legs,
one for the component and one for the sites.
In this matrix notation the Hamiltonian reads
\begin{equation}\label{eq:hamiltonian_beb-form}
  \hat{H}
  = \sum_{\sigma} \mat{\oc}^{\dagger}_{\sigma}\cmp{\eta}^{\transp} \cmp{H}_{\sigma} \cmp{\eta}\mat{\oc}_{\sigma}
  + \mat{\hat{n}}^{\transp}_{\uparrow} \cmp{\eta}^{\transp} \cmp{U} \cmp{\eta}  \mat{\hat{n}}_{\downarrow},
\end{equation}
where we introduced the local interaction tensor \({(\cmp{U})}^{\alpha \beta}_{ij} = \delta_{ij} \cmp*{U}^{\alpha} \delta^{\alpha \beta}\).
In the non-interacting case the Hamiltonian matrix of a specific configuration is the extended matrix sandwiched by the indicator tensors
\begin{equation}\label{eq:extendedH}
  \mat{H}
  = \cmp{\eta}^{\transp} \cmp{H} \cmp{\eta}
  =
  \begin{tikzpicture}[baseline=(eta.base), minimum size=\nodesize, node distance=0.5cm]
    \node[shape=rectangle, draw] (etaT) at (0, 0) {\(\cmp{\eta}^{\transp}\)};
    \node[shape=rectangle, draw] (cmptH) [right=of etaT] {\(\cmp{H}\)};
    \node[shape=rectangle, draw] (eta) [right=of cmptH] {\(\cmp{\eta}\)};
    \draw ([yshift=-1ex]etaT.west) -- ++ (-\olegsize, 0);
    \draw ([yshift=1ex]etaT.east) -- ([yshift=1ex]cmptH.west) ([yshift=1ex]cmptH.east) -- ([yshift=1ex]eta.west);
    \draw ([yshift=-1ex]etaT.east) -- ([yshift=-1ex]cmptH.west) ([yshift=-1ex]cmptH.east) -- ([yshift=-1ex]eta.west);
    \draw (eta.-30) -- ++ (\olegsize, 0);
  \end{tikzpicture}.
\end{equation}
For every lattice site in the extended representation each component of the non-interacting Hamiltonian matrix \(\cmp{H}\) is assigned a corresponding element.
In this way the non-interacting Hamiltonian can be generated by \(\cmp{H}\) for every disorder configuration.
The matrix product in the algebraic equation equals the tensor contractions of the internal legs as illustrated by the right hand side of \cref{eq:extendedH}.
\Cref{sec:example} provides an explicit example for a system of \(N=3\) sites and \(M=2\) components.
In \cref{tab:disorder-symbols} we collect the symbols used in our paper.
We note, that the only configuration dependent parts in \cref{eq:hamiltonian_beb-form} are the matrices \(\cmp{\eta}\) and \(\cmp{\eta}^{\transp}\);
the rest is independent of the specific disorder realization.
In other words:
Comparing \cref{eq:hamiltonian_beb-form} with \cref{eq:hamiltonian_mat-form}
the configuration dependence of the former equation is moved from the Hamiltonian matrix
to the local indicator tensors \cref{eq:indicator-tensor}.
This is the main point of the BEB algorithm:
One can work with a non-random but extended Hamiltonian matrix \(\cmp{H}\),
which contains the parameters for all possible configurations.
A specific configuration can be selected by applying indicator tensors \(\cmp{\eta}\).
What remains to be averaged over are these local indicator tensors.

\subsubsection{Alloy component Green's function}
In the absence of interaction, \(\cmp{U} = 0\),
the model can be solved by the generalized CPA introduced by Blackman, Esterling and Berk~\cite{Blackman_1971,Esterling_1975} (BEB).
Using the indicator tensor \(\cmp{\eta}\) \cref{eq:indicator-tensor}
we define the projector:
\begin{equation}\label{eq:def:projector}
  \cmp{\chi}
  = \cmp{\eta \eta}^{\transp}
  =
  \begin{tikzpicture}[baseline=(eta.base), minimum size=\nodesize, node distance=0.5cm]
    \node[rectangle, draw] (eta) at (0, 0) {\(\cmp{\eta}\)};
    \node[rectangle, draw] (etaT) [right=of eta] {\(\cmp{\eta}^{\transp}\)};
    \draw ([yshift=-1ex]eta.west) -- ++ (-\olegsize, 0);
    \draw ([yshift=+1ex]eta.west) -- ++ (-\olegsize, 0);
    \draw ([yshift=-1ex]eta.east) -- ([yshift=-1ex]etaT.west);
    \draw ([yshift=-1ex]etaT.east) -- ++ (\olegsize, 0);
    \draw ([yshift=+1ex]etaT.east) -- ++ (\olegsize, 0);
  \end{tikzpicture};
  \qquad%%
  \cmp{\chi}^{2}
  = \cmp{\chi}.
\end{equation}
It maps a vector in the extended space onto a single configuration;
all elements corresponding to different configurations are set to \(0\).
The projector property follows from the indicator identity \(\cmp{\eta}^{\transp} \cmp{\eta} = \I\).
For the non-interacting system, we define the component Green's function as
\begin{equation}\label{eq:cmpG}
  \begin{aligned}
  &\cmp{G}(z)
  \coloneqq \cmp{\eta} \mat{G}(z) \cmp{\eta}^{\transp}
  =
  \begin{tikzpicture}[baseline=(eta.base), minimum size=\nodesize, node distance=0.5cm]
    \node[rectangle, draw] (eta) at (0, 0) {\(\cmp{\eta}\)};
    \node[rectangle, draw] (Gf) [right=of eta] {\(\mat{G}(z)\)};
    \node[rectangle, draw] (etaT) [right=of Gf] {\(\cmp{\eta}^{\transp}\)};
    \draw ([yshift=-1ex]eta.west) -- ++ (-\olegsize, 0);
    \draw ([yshift=+1ex]eta.west) -- ++ (-\olegsize, 0);
    \draw ([yshift=-1ex]eta.east) -- ([yshift=-1ex]Gf.west);
    \draw ([yshift=-1ex]Gf.east) -- ([yshift=-1ex]etaT.west);
    \draw ([yshift=-1ex]etaT.east) -- ++ (\olegsize, 0);
    \draw ([yshift=+1ex]etaT.east) -- ++ (\olegsize, 0);
  \end{tikzpicture};
  \\%%
  &\sum_{\alpha \beta} \cmp{G}^{\alpha \beta}(z)
  = \mat{G}(z).
\end{aligned}
\end{equation}
The arrangement of indicator tensors \(\cmp{\eta}\) is different compared to \cref{eq:extendedH}:
Both the Green's function \(\mat{G}\) and the component Green's function \(\cmp{G}\) are configuration dependent.
We note that local elements are diagonal in component space, i.e., \(\cmp*{G}^{\alpha \beta}_{ii}(z) \propto \delta^{\alpha \beta}\).
We sandwich the resolvent for the Green's function
\begin{equation}
  \I = [\I z - \mat{H}]\mat{G}(z)
\end{equation}
by \(\cmp{\eta}\) from the left and \(\cmp{\eta}^{\transp}\) from the right;
this yields the equation for the component Green's function
\begin{equation}
  \cmp{\chi}
  = [\I z - \cmp{\chi} \cmp{H} \cmp{\chi}] \cmp{G}(z).
\end{equation}
The law of total probability~\cite{ka.ta.75} relates the average of the component Green's functions
and the conditional average of the physical Green's function in the following way:
\begin{equation}\label{eq:cmpG_physG}
  \Expval(\cmp*{G}^{\alpha \beta}_{ij})
  = \begin{cases}{}
    \phantom{c^{\beta}} c^{\alpha} \Expval(G_{ii}(z)\vert i \mapsto \alpha) \delta^{\alpha \beta}
    & \qfor i = j,
    \\
    c^{\alpha} c^{\beta} \Expval(G_{ij}(z)\vert i,j \mapsto \alpha, \beta)
    & \qfor i \neq j.
  \end{cases}
\end{equation}

\subsubsection{Effective medium in the extended space}
As in CPA,
in the BEB formalism one calculates an effective local Green's function \(\Gloc(z)\) from an effective medium \(\cmp{S}(z)\),
which approximates the average local Green's function \(\Expval(\cmp{G}_{ii}(z))\).
We consider only substitutional disorder without structural disorder, i.e.,
the lattice structure is assumed to be fixed.
Therefore, we decompose the hopping tensor \(\cmp*{t}^{\alpha \beta}(\abs{\vb{r}_{i} - \vb{r}_{j}})\) into its component part, \(\cmp*{T}^{\alpha \beta}\), and its lattice part, \(t(\abs{\vb{r}_{i} - \vb{r}_{j}})\):
\begin{equation}
  \cmp*{t}^{\alpha \beta}(\abs{\vb{r}_{i} - \vb{r}_{j}})
  \eqqcolon \cmpT*^{\alpha \beta}\, t(\abs{\vb{r}_{i} - \vb{r}_{j}}).
\end{equation}
Depending on the component of the endpoints,
the matrix elements \(\cmpT*^{\alpha \beta}\) scale the amplitudes for hopping on a given lattice structure by a dimensionless factor.
In the following we refer to \(\cmpT*^{\alpha \beta}\) simply as ``dimensionless hopping parameter''.
We perform the lattice Fourier transform for the hopping matrix elements as
\begin{equation}
  \cmpT \frac{1}{N} \sum_{ij} t(\abs{\vb{r}_{i} - \vb{r}_{j}}) e^{i\vb{k}\cdot(\vb{r}_{i} - \vb{r}_{j})}
  = \cmpT \epsilon_{k}.
\end{equation}
For a given effective medium \(\cmp{S}(z)\),
the effective local Green's function reads
\begin{equation}\label{eq:BEB_Gloc}
  \Gloc(z)
  = \frac{1}{N} \sum_{k} {[\I z - \cmp{S}(z) - \cmpT \epsilon_{k}]}^{-1}.
\end{equation}
The effective medium as well as the effective local Green's function are represented by \(M \times M\) matrices in the components.
Being local quantities they no longer carry lattice indices.
The effective medium \(\cmp{S}(z)\) is determined by demanding that the averaged \(\tmat\)-matrix vanishes:
\begin{gather}
  \Expval(\tmat(z))
  \overset{!}{=} 0, \label{eq:avg_tmat0}
  \\%%
  \tmat(z)
  \coloneqq -{[
    \cmp{\chi}{[\I - \cmp{\chi}(\I z - \cmp{S}(z) + \cmp{v})\cmp{\chi}]}^{-1} \cmp{\chi}
    + \Gloc(z)
  ]}^{-1}.\label{eq:def:tmat}
\end{gather}

\subsection{Inclusion of electronic interactions and the BEB+DMFT self-consistency loop}\label{sec:self}
We treat the local Hubbard interaction within the dynamical mean-field theory~\cite{me.vo.89,ge.ko.96,ko.vo.04},
which assumes a local self-energy \(\Sigma_{ij}(z) = \delta_{ij} \Sigma_{ii}(z)\);
this property becomes exact in the limit of infinite coordination number.
The problem of interacting disordered electrons may equally be viewed as a system of non-interacting particles moving in an effective local, energy dependent potential $\Sigma_{ii}(z)$; for details see Ref.~\onlinecite{ja.vo.92,Vollhardt_2010}.
The DMFT self-consistency equations~\cite{ge.ko.96} are equivalent to a fixed-point problem
which can be expressed by a functional \(\widehat{\Sigma}\):
Given a self-energy \(\Sigma_{ii}\) and the resulting local Green's function \(G_{ii}(\Sigma_{ii})\)
this functional provides a new self-energy \(\widehat{\Sigma}\big[G_{ii}(\Sigma_{ii}), \Sigma_{ii}\big]\),
such that the DMFT self-energy is determined self-consistently by the fixed-point
\begin{equation}
  \Sigma_{ii}
  = \widehat{\Sigma}\big[G_{ii}(\Sigma_{ii}), \Sigma_{ii}\big].
\end{equation}
Within the CPA,
the local Green's function for a given self-energy \(G_{ii}(\Sigma_{ii})\) is replaced by the conditional average \(\Expval(G_{ii}(\Sigma_{ii})|i\mapsto\alpha)=\Gloc*^{\alpha \alpha}(\Sigma_{ii})/c^\alpha\), see \cref{eq:cmpG_physG,eq:BEB_Gloc}.
Thus, the self-energy \(\widehat{\Sigma}\big[\Gloc*^{\alpha\alpha}(\Sigma_{ii})/c^{\alpha}, \Sigma_{ii}\big]\) depends on the component \(\alpha\).
Consequently, the self-energy at the fixed-point depends on the component \(\alpha\), but not on the explicit site \(i\):
\begin{equation}\label{eq:DMFT_component-fixed-point}
  \Sigma^{\alpha}
  =\widehat{\Sigma}\big[\Gloc*^{\alpha\alpha}(\Sigma^{\alpha})/c^\alpha, \Sigma^{\alpha}\big].
\end{equation}
This allows one to introduce the BEB+DMFT self-consistency
which we will discuss next.
By merging the BEB formalism with DMFT a two-fold self-consistency arises,
one for the BEB and one for the DMFT corresponding to the fixed-point \cref{eq:DMFT_component-fixed-point}.
The self-consistency equation of the BEB formalism is pointwise in the frequencies and is therefore much simpler than the self-consistency condition of the DMFT,
where frequencies mix due to the energy exchange caused by the interaction between the electrons.
We view the former self-consistence as an internal part of the full self-consistency loop.
In the BEB method we calculate an effective local Green's function \(\Gloc(z)\), \cref{eq:BEB_Gloc}.
The effective medium \(\cmp{S}(z)\) and, therefore, the effective local Green's function have to be calculate self-consistently from \cref{eq:avg_tmat0,eq:def:tmat}.
This condition simplifies to
\begin{equation}\label{eq:BEB_self-consistency}
  \Gloc^{-1}(z) = \overline{\cmp{g}}^{-1}(z)
\end{equation}
with the diagonal matrix
\begin{equation}\label{eq:ave}
  {\overline{\cmp*{g}}}^{\alpha \beta}(z)
  = \frac{c^{\alpha} \delta^{\alpha \beta}}{{(\Gloc^{-1})}^{\alpha \alpha}(z) + \cmp*{S}^{\alpha \alpha}(z) + \mu - \cmp*{v}^{\alpha} - \Sigma^{\alpha}(z)},
\end{equation}
where \(\Sigma^{\alpha}(z)\) is the DMFT self-energy for the component \(\alpha\).
The self-consistent \cref{eq:BEB_self-consistency} can be solved with standard root-search algorithms
or by simple iteration.
In practice, we use an implementation of the BEB formalism without interactions and merely shift the on-site energy \(\cmp*{v}^{\alpha} \rightarrow \cmp*{v}^{\alpha} + \Sigma^{\alpha}(z)\).
An efficient evaluation of the BEB self-consistency equation is discussed in \cref{sec:BEB-Gf_implementation};
an implementation is provided in Ref.~\onlinecite{we.oe.21}.
To emphasize the dependence on the self-energy
we denote the self-consistently determined effective local Green's function for a given self-energy \cref{eq:BEB_self-consistency} by \(\Gloc(z, \mat{\Sigma}(z))\).
With the BEB self-consistency condition \cref{eq:BEB_self-consistency} for the local Green's function \(\Gloc(z, \mat{\Sigma}(z))\),
the combined algorithm corresponds to the conventional DMFT self-consistency condition
\cref{eq:DMFT_component-fixed-point},
where the local Green's function,
calculated from the lattice Hilbert transform,
is now replaced by the average
\begin{equation*}
  \Expval(G_{ii}\vert i \mapsto \alpha) = \Gloc*^{\alpha \alpha}(z, \mat{\Sigma}(z)) / c^{\alpha}.
\end{equation*}
The reciprocal concentration factor can be avoided by introducing a renormalized indicator tensor,
which leads to a slightly modified BEB self-consistency as elaborated in \cref{sec:renormalized-BEB}.
We have to solve a separate impurity problem for every component \(\alpha\).
Starting from an initial guess for the DMFT self-energy \(\Sigma^{\alpha}(z)\) for every component,
the BEB+DMFT scheme is the following:
\begin{subequations}
\begin{enumerate}
  \item Calculate the effective local Green's function \cref{eq:BEB_Gloc} using \cref{eq:BEB_self-consistency,eq:ave}, which yields
    \begin{equation}
      \Gloc(z, \mat{\Sigma}(z)),
    \end{equation}
  \item calculate the hybridization function
    \begin{equation}\label{eq:impurity_hybridization}
      \Delta^{\alpha}(z) = z + \mu - \cmp*{v}^{\alpha} - \Sigma^{\alpha}(z) - c^{\alpha}/\Gloc*^{\alpha \alpha}(z, \mat{\Sigma}(z)),
    \end{equation}
    for every component $\alpha$,
  \item solve the impurity problem for the self-energy
    \begin{equation}\label{eq:impurity_self-energy}
      \Sigma^{\alpha}(z) = \Sigma[\cmp*{v}^{\alpha}, \cmp*{U}^{\alpha}, \Delta^{\alpha}]
    \end{equation}
    for every component $\alpha$,
  \item repeat from step 1 until self-consistency is reached.
\end{enumerate}
\end{subequations}
The hybridization function can also be expressed in terms of BEB quantities using the self-consistency condition \cref{eq:BEB_self-consistency}:
\begin{equation}\label{eq:hyb_BEB-DMFT}
  \Delta^{\alpha}(z)
  = z - \cmp*{S}^{\alpha \alpha}(z) - {(\Gloc^{-1})}^{\alpha \alpha}(z, \mat{\Sigma}(z)).
\end{equation}
This is different from CPA+DMFT,
where only one unique hybridization functions exists independent of the alloy components.
Analogous to the non-disordered case,
an expression for the hybridization function of the Bethe lattice in terms of the local Green's function \(\Gloc(z, \mat{\Sigma}(z))\) is given in \cref{sec:Bethe_hybridization}.
Central to the DMFT problem is the impurity solver which provides the local dynamic self-energy \cref{eq:impurity_self-energy}.
To this end, we employ a tensor network based zero temperature solver,
the fork tensor-product state (FTPS) solver\cite{ba.zi.17}.
The FTPS impurity solver is a Hamiltonian-based method
which discretizes the hybridization function \cref{eq:impurity_hybridization} using a large number of bath sites.
We use \(249\) sites per spin resulting in a median energy distance of \(0.03D\),
where the half-bandwidth \(D\) sets our energy scale.
We calculate the ground state \(\ket{GS}\) of the finite size impurity problem using the density matrix renormalization group (DMRG)\cite{whit.92,scho.05}.
Subsequently, we perform the time evolution using the time dependent variational principle (TDVP)\cite{ha.ci.11,lu.os.15,ha.lu.16,ba.ai.20}.
To obtain the retarded time impurity Green's function \(\Gtret(t)\),
the states \(\oc_{\sigma}\ket{GS}\), \(\ocd_{\sigma}\ket{GS}\), as well as their adjoint states are time-evolved,
where \(\oc_{\sigma}\) (\(\ocd_{\sigma}\)) is the annihilation (creation) operator of the impurity site.
For DMRG we chose a truncated weight of \(10^{-15}\) and a maximal bond-dimension of \(100\).
We perform the TDVP using time-steps of \(0.1/D\) up to a maximal time \(t_{\text{max}} = 150/D\)
with a truncated weight of \(10^{-9}\) and a maximal bond-dimension of \(150\).
The convergence with respect to these parameters is checked.
We can calculate the Green's function \(\Gtret(t)\) only up to a maximal time,
and we have (small) finite size effects due to the discretization of the bath.
Therefore, we cannot evaluate the retarded Green's function directly on the real-frequency axis \(G(\omega + i0^+)\). 
Instead we calculate it on a parallel contour \(G(\omega + i\eta)\) shifted by a fixed finite \(\eta > 0\);
this corresponds to the Laplace transform:
\begin{equation}
  G(\omega + i \eta)
  = \int_{0}^{\infty} dt \e^{i(\omega + i\eta)t} \Gtret(t)
  \eqqcolon \Fourier_{\eta}[\Gtret(t)](\omega).
\end{equation}
The shift \(\eta\) acts as a broadening for the Green's function \(G(z)\) as can be seen from the Cauchy integral formula
\begin{equation}
  2 \pi iG(\omega + i \eta)
  = \oint\! dz \frac{G(z)}{z - \omega - i \eta}
  = \int_{-\infty}^{\infty}\!\! d \omega^{\prime} \frac{G( \omega^{\prime})}{\omega^{\prime} - \omega - i \eta^{-}},
\end{equation}
with \(\eta^{-} = \eta - 0^{+}\), where \(0^{+}\) is a positive infinitesimal.
We write the Green's function on the real axis in terms of the shifted Fourier transform
\begin{equation}
\begin{aligned}
  G(\omega + i0^+)
  &= \lim_{\eta^{\prime} \searrow 0} \Fourier_{{\eta}^{\prime}}[\Gtret(t)](\omega)
  = \Fourier_{\eta}[\e^{t \eta^{-}}\Gtret(t)](\omega)
  \\%%
  &= \sum_{k} \frac{1}{k!} {(\eta^{-})}^{k} \Fourier_\eta[t^{k}\Gtret(t)](\omega).
\end{aligned}
\end{equation}
The second equality replaces the limit \({\eta}^{\prime}\searrow 0\) using \(\lim_{{\eta}^{\prime} \searrow 0} \exp(-{\eta}^{\prime} t) = \exp(-\eta t)\exp(\eta^{-} t)\) introducing a finite variable \(\eta\),
in the last line we use the series representation of the exponential function \(\exp(\eta^{-} t)\).
The first term \(k=0\) is the Green's function on the shifted contour \(G(\omega + i \eta)\),
higher order terms give systematic corrections.
In \cref{sec:results}, we calculate the first order correction
\begin{equation}
  G(\omega + i0^{+})
  = G(\omega + i\eta) + \eta \Fourier_\eta\left[t\Gtret(t)\right](\omega) + \order{\eta^{2}}
\end{equation}
with a typical shift \(\eta = 0.08\).
The self-energy is calculated from the equation of motion of the impurity model~\cite{bu.he.98}
\begin{align}
  \Sigma_\sigma(z)
  &= UF_{\sigma}(z)/G_{\sigma}(z),
  \\%%
  {F}^{\text{ret}}(t)
  &= \expval{\oc_{\sigma}(t)\hat{n}_{- \sigma}(t) \ocd_{\sigma}}{GS},
\end{align}
where \(F(z)\) is the Laplace transform of \({F}^{\text{ret}}(t)\).

\subsection{General properties of the BEB formalism}
We shortly review some properties of the BEB formalism~\cite{go.ga.77,Gonis_1992,ko.ve.98}.
First, the BEB formalism
is equivalent to the CPA when off-diagonal disorder is absent.
This limit was already proven in the original formulation~\cite{Blackman_1971,Esterling_1975}.
Since the BEB theory includes the off-diagonal disorder in the single-site approximation,
the Herglotz property\cite{mull.73,mi.ra.78} of the CPA is preserved in the BEB as well~\cite{go.ga.77}.
Second, for a non-interacting tight-binding Hamiltonian
the density of states (DOS) is non-zero only within certain energy ranges, determined by the Hamiltonian matrix elements.
This is also holds for the CPA~\cite{ve.ki.68}.
Koepernick \emph{et al.}~\cite{ko.ve.98} found the same
in their numerical study of one-dimensional chains using the BEB formalism.
Likewise, we find no violations of this property in our numerical results.
In the following subsection we derive exact bounds for the spectral function of a Bethe lattice with coordination number \(Z\) using the BEB formalism in the limit of independent alloy components, i.e., when there is no hopping between different components.

\subsection{Limit of independent components}\label{sec:bandwidth_scaling}
We consider the limit of vanishing hopping between different components.
If the hopping is diagonal in the components, \(\cmpT*^{\alpha \beta} \propto \delta^{\alpha \beta}\),
the BEB effective medium \(\cmp{S}(z)\) is also diagonal in the components
and the self-consistency equations \cref{eq:BEB_self-consistency} decouple.
In this case, the effective local Green's function \cref{eq:BEB_Gloc} can be readily calculated,
since the matrix inverse is the reciprocal of the diagonal elements
\begin{equation}
\begin{aligned}
  \Gloc*^{\alpha \beta}(z)
  &= \frac{1}{N} \sum_{k} \frac{\delta^{\alpha \beta}}{z - \cmp*{S}^{\alpha \alpha}(z) - \cmpT*^{\alpha \alpha} \epsilon_{k}}
  \\%%
  &= \delta^{\alpha \beta} g_{0}^{\alpha}\big(z - \cmp*{S}^{\alpha \alpha}(z)\big).
\end{aligned}
\end{equation}
Here, \(g_{0}^{\alpha}\) is the lattice Hilbert transform \(g_{0}(z) = \frac{1}{N} \sum_{k} \frac{1}{z - \epsilon_{k}}\);
the superscript \(\alpha\) indicates that the bandwidth is scaled by \(\cmpT*^{\alpha \alpha}\):
\begin{equation}
  g_{0}^{\alpha}(z)
  = \frac{1}{N} \sum_{k} \frac{1}{z - \cmpT*^{\alpha \alpha} \epsilon_{k}}
  = \frac{1}{\cmpT*^{\alpha \alpha}} g_{0}(z/\cmpT*^{\alpha \alpha}).
\end{equation}
For the component \(\alpha\),
the decoupled self-consistency \cref{eq:BEB_self-consistency} reads
\begin{equation}
  0
  = \frac{\overline{c}^{\alpha}}{g_{0}^{\alpha}(z - \cmp*{S}^{\alpha \alpha})} + \cmp*{S}^{\alpha \alpha} - \cmp*{v}^{\alpha},
\end{equation}
with the concentration complement \(\overline{c}^{\alpha} = 1 - c^{\alpha} \geq 0\).
For a Bethe lattice with coordination number \(Z\) and the lattice Hilbert transform\cite{econ.06}
\begin{align} % following `widetext` has trouble with equation, cmp. https://tex.stackexchange.com/a/120384/146727
  g_{0}(z, Z)
  = 2(Z-2) /\left[z \left(Z - 2 + Z\sqrt{1 - D^{2}/ z^{2}}\right)\right],
\end{align}
where \(D\) is the half-bandwidth,
the self-consistency condition is an algebraic equation and can be solved analytically.
The BEB effective medium reads
\begin{widetext}
\begin{equation}
  \cmp*{S}^{\alpha \alpha}(z, Z)
  = \frac{(Z-2)\cmp*{v}^{\alpha} c^{\alpha} + Z \cmp*{v}^{\alpha} + (Z+2)c^{\alpha} z - Zz - 2{(c^{\alpha})}^2z
          - Z \overline{c}^{\alpha} s\sqrt{{(z-\cmp*{v}^{\alpha})}^{2} - c^{\alpha} {(D^{\alpha})}^{2} \frac{Z-c^{\alpha}}{Z-1}}}
    {2c^{\alpha}(Z-c^{\alpha})},
\end{equation}
\end{widetext}
where \(s\) is the sign \(s=\sign(\Re(z - v^{\alpha}))\),
and \(D^{\alpha}\) is the half-bandwidth scaled by \(\cmpT*^{\alpha \alpha}\);
this is the retarded solution.
A conjugate solution exists with \(-s\) and therefore with a plus sign in front of the square root.
We are interested in the bandwidth of the resulting component spectrum
\begin{equation}
  A^{\alpha}(\omega)
  = - \frac{1}{c^{\alpha}\pi} \Im g_{0}^{\alpha}(\omega + i0^{+} - \cmp*{S}^{\alpha \alpha}(\omega + i0^{+})).
\end{equation}
For non-interacting systems,
the Gershgorin circle theorem\cite{gers.31} gives the \emph{maximal} spectral bounds
\begin{equation}
  \abs{z - \cmp*{v}^{\alpha}} \leq D^{\alpha}.
\end{equation}
In the limit \(\cmpT*^{\alpha \beta} \propto \delta^{\alpha \beta}\),
we can make a more precise statement and derive exact spectral bounds as will be discussed below.
The spectral function can only vanish when the imaginary part of the effective medium vanishes.
Thus, for non-interacting systems,
we need to check where the argument of the square root is negative.
One finds an imaginary part and therefore spectral weight for
\begin{equation}
  \abs{z - \cmp*{v}^{\alpha}}
  < \sqrt{c^{\alpha} \frac{Z - c^{\alpha}}{Z - 1}} D^{\alpha}.
\end{equation}
Therefore, for the Bethe lattice with coordination number \(Z\) and \(\cmpT*^{\alpha \beta} \propto \delta^{\alpha \beta}\),
the bandwidth is reduced due to concentration by a factor \(\sqrt{c^{\alpha} (Z - c^{\alpha})/(Z - 1)}\).
We then obtain the effective bandwidth
\begin{equation}\label{eq:bandwidth-scaling_Bethe}
  D^{\alpha}_{\text{eff}}
  = \sqrt{c^{\alpha} \frac{Z - c^{\alpha}}{Z - 1}} \cmpT*^{\alpha \alpha} D.
\end{equation}
Our numerical results in \cref{sec:results} were obtained for a semicircular DOS, i.e.,
the Bethe lattice with infinite coordination number \(Z \rightarrow \infty\).
In this limit one finds an effective bandwidth
\begin{equation}
  D^{\alpha}_{\text{eff}}
  = \sqrt{c^{\alpha}} \cmpT*^{\alpha \alpha} D.
\end{equation}
The same factor \(\sqrt{c}\) was found in Ref.~\onlinecite{kr.ho.04} in the CPA (\(\tAA = \tAB = \tBB = 1\))
in the limit of high disorder strength \((\vB - \vA)/D = \delta \gg \max(1, U/D)\).
While the parameters in these limits are different, both describe the same physics, namely the decoupling of components.
Indeed, the components decouple not only for
vanishing hopping between the components \(\tAB = 0\),
but also in the case of a large separation in energy (\(\delta \gg 1\)).
For coordination number \(Z=2\) another interesting limit of the Bethe lattice is obtained;
this is the one-dimensional lattice\cite{econ.06}, where
\begin{equation}
  g^{1\text{D}}_{0}(z)
  = g_{0}(z, Z=2)
  = 1 / \left[z \sqrt{1 - D^2/z^2}\right].
\end{equation}
The spectral bounds are given by
% checked for c=0.5
\begin{equation}
  D^{\alpha}_{\text{eff}}
  = \sqrt{c^{\alpha}(2 - c^{\alpha})} \cmpT*^{\alpha \alpha} D.
\end{equation}
Therefore, for the one-dimensional lattice and \(\cmpT*^{\alpha \beta} \propto \delta^{\alpha \beta}\) the bandwidth is reduced by the factor \(\sqrt{c^{\alpha}(2 - c^{\alpha})}\).

\section{Numerical results}\label{sec:results}
The above formalism is now used to study the effect of off-diagonal disorder in the Anderson--Hubbard model at zero temperature.
We employ a Bethe lattice with infinite coordination number, whose half-bandwidth \(D\) sets the energy scale.
Furthermore we consider a discrete binary random alloy distribution with components \(\CmpA\) and \(\CmpB\).
In all applications we consider the case of half-filling on average\(\Expval(n_{i})=1\);
this leads to a fixed chemical potential which we choose as \(\mu=0\).
In the following subsection we fix the alloy component concentration and study the change in the spectral function starting with the non-interacting case and equal atomic potentials.
The alloy component spectral functions are the concentration-weighted conditional spectral functions
\begin{equation}
  \cmp*{A}^{\alpha}(\omega)
  \coloneqq - \frac{1}{\pi} \Im \Gloc*^{\alpha \alpha}(\omega)
  = - \frac{c^{\alpha}}{\pi} \Im \Expval(G_{ii}(\omega)|i \mapsto \alpha).
\end{equation}
The average spectral functions are given by the trace
\begin{equation}
  \Expval(A(\omega))
  = \sum_{\alpha} \cmp*{A}^{\alpha}(\omega)
  = - \frac{1}{\pi} \Im \Tr \Gloc(\omega).
\end{equation}

\subsection{Non-interacting limit}\label{subsec:non-interacting}
We start with the non-interacting case by setting \(\UA = \UB = 0\),
which corresponds to the Anderson disorder model with purely off-diagonal disorder.
Since the non-interacting Green's function is independent of temperature 
the results presented in this section are valid not only for zero, but also for finite temperatures.
We choose the parameters
\begin{equation*}
  \vA = \vB = -U/2 = 0;~%
  c^\CmpA = 0.1 = 1 - c^\CmpB;~%
  \tAA = \tBB = 1
\end{equation*}
and calculate the average and the alloy component spectral functions for several values of \(\tAB\) at half filling.
The case \(\tAB=1\) is equivalent to the non-disordered case since \(\vA=\vB\);
in this case the components are indistinguishable.
Thus, the average spectral function is just the spectral function of the non-disordered Bethe lattice,
and the component Green's function are proportional with a concentration prefactor.
\begin{figure}[htb]
  \includegraphics[width=\linewidth]{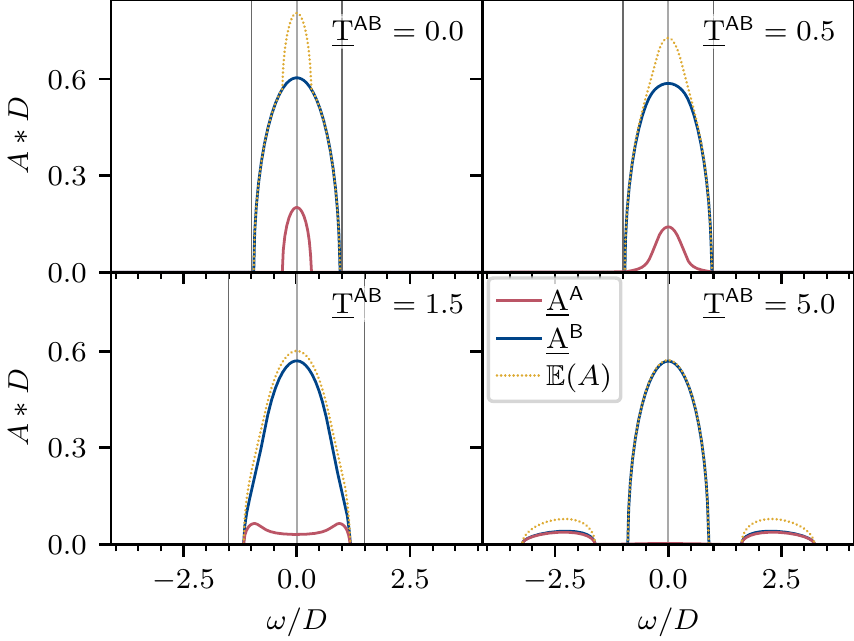}
  \caption{%
    Non-interacting case: Comparison of spectral functions for different dimensionless hopping parameters \(\tAB\).
    The parameters are \(\UA=\UB=0\), \(\vA=\vB=0\), \(\cA=0.1=1-\cB\), \(\tAA=\tBB=1\).
    The solid lines represent the component spectral functions \(\cmp*{A}^{\alpha}(z) = c^{\alpha} \Expval(A_{i}(z) \vert i \mapsto \alpha)\),
    where \(\CmpA\) is red and \(\CmpB\) is blue;
    the dotted yellow line shows the average spectral function \(\Expval(A(z)) = \cmp*{A}^{\CmpA}(z) + \cmp*{A}^{\CmpB}(z)\).
    The thin vertical lines show the maximal spectral bounds given by the Gershgorin circle theorem\cite{gers.31}.
  }\label{fig:non_int}
\end{figure}
\Cref{fig:non_int} shows the spectral function for off-diagonal disorder with \(\tAB = 0.0, 0.5, 1.5, 5.0\).
The case \(\tAB=0\) was solved exactly in \cref{sec:bandwidth_scaling}.
The panel \(\tAB = 0\) in \cref{fig:non_int} indicates that off-diagonal disorder reduces the bandwidths;
according to \cref{eq:bandwidth-scaling_Bethe} the effective bandwidths are given by \(D_{\text{eff}}^\CmpA = \sqrt{0.1}D \approx 0.32 D\) and \(D_{\text{eff}}^\CmpB = \sqrt{0.9}D \approx 0.95D\).
For \(\tAB < 1 = \cmpT*^{\alpha \alpha}\) the probabilities for hopping between the alloy components \(\CmpA\) and \(\CmpB\) are less than those between the same component \(\alpha\).
The spectral functions in the upper half of~\cref{fig:non_int} correspond to this situation.
In spite of a similar support on the energy axis, the spectral function of the majority component \(\CmpB\) has a larger bandwidth, which encompasses the effective bandwidth of component \(\CmpA\).
By contrast, when \(\tAB>1=\cmpT*^{\alpha \alpha}\), \(\CmpA\)-\(\CmpB\) bonds are energetically favorable.
The panel \(\tAB=1.5\) in \cref{fig:non_int} shows that 
the spectral function of component \(\CmpA\) develops shoulders although both components have similar effective bandwidths.
When the value of \(\tAB\) is increased further the shoulders split off from the central band;
for the parameters chosen the split-off is visible for \(\tAB \geq 2.25\).
According to Burdin and Fulde~\cite{Burdin_2007}
the split-off upper and lower bands correspond to bonding and antibonding states, respectively.
In the particle-hole symmetric case,
the bonding and antibonding bands in panel \(\tAB=5.0\) of \cref{fig:non_int} have equal weights.
Due to the large value \(\tAB = 5.0\), the minority component \(\CmpA\) is completely suppressed in the central band.
The components \(\CmpA\) and \(\CmpB\) contribute roughly equally to the bonding and antibonding subbands.
Therefore the central band for the \(\CmpB\) component is depleted by an amount of \((1-2c_A)=0.8\) of the spectral function.
The overall results and the spectral weight transfer from the central band are consistent with those reported in Ref.~\onlinecite{Burdin_2007}.

\subsection{Alloy components with equal interaction strengths \texorpdfstring{\(\UA=\UB\)}{U\^A=U\^B}}
In the following, we will discuss the results for the interacting case using the setup described in \cref{subsec:non-interacting} at zero temperature ($T=0$).
The alloy components have identical on-site interaction parameters:
\begin{equation*}
  \UA = \UB = U = 3D.
\end{equation*}
At half-filling, the on-site energies are \(\vA = \vB = -U/2 = -1.5 D\).
\begin{figure}[htb]
  \includegraphics[width=\linewidth]{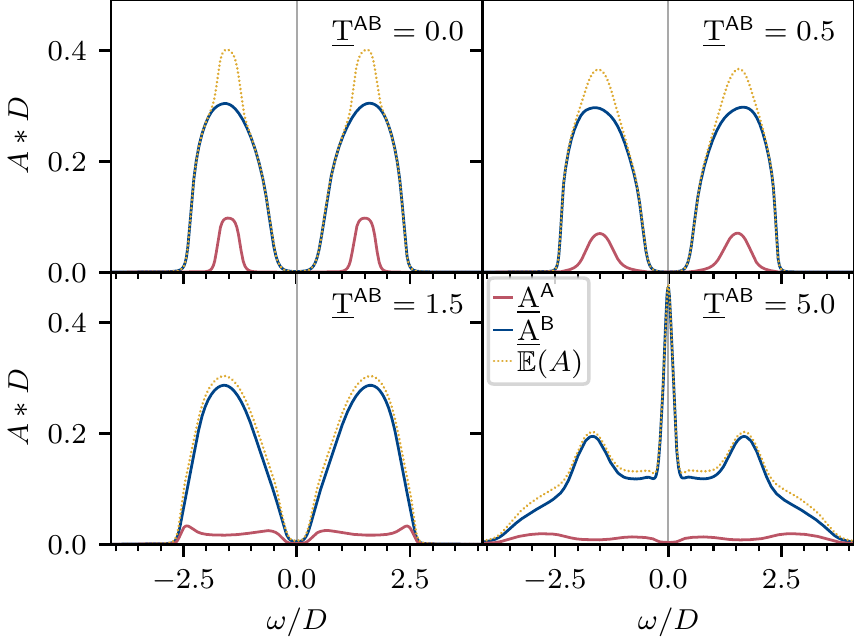}
  \caption{%
    Comparison of spectral functions for different values of the dimensionless hopping parameter \(\tAB\) with
    \(\UA = \UB = 3 D\), \(\vA=\vB=-1.5 D\), \(\cA = 0.1 = 1 - \cB\), and \(\tAA=\tBB=1\) at \(T=0\).
    For \(\tAB = 0\) a shift \(\eta = 0.12\) had to be used;
    the other panels were calculated for \(\eta = 0.08\).
  }\label{fig:sameU}
\end{figure}
\Cref{fig:sameU} shows the spectral function for various values of \(\tAB\).
For \(\tAB=0\), when the self-consistency equations decouple,
the spectral function of both components implies insulating behavior due to the strong interaction \(U=3D\).
The upper and lower Hubbard bands centered around \(\pm U/2\) are visible.
The bandwidth of the Hubbard bands is effectively reduced according to \cref{eq:bandwidth-scaling_Bethe}.
Therefore the gap is wider for component \(\CmpA\).
Finite values of \(\tAB\) lead to wider spectral functions,
and the bandwidth of the minority component \(\CmpA\) broadens to the same bandwidth as for \(\CmpB\).
Although for \(\tAB = 1.5\) the imaginary part of the self-energy shows a prominent peak at \(\omega = 0\),
the spectral function remains finite at the Fermi level.
The convergence of the DMFT computations slows down for this value of \(\tAB=1.5\), hinting at the proximity of a transition.
The minority component \(\CmpA\) exhibits shoulders at the band-edge.
For \(\tAB=1.7\) (not shown) one observes a pronounced quasi-particle peak at the Fermi level of both components, indicating that the system is metallic.
A further increase to \(\tAB=5.0\) leads to an increased spectral weight at the Fermi level for the majority component \(\CmpB\),
while the spectral function of \(\CmpA\) has a minimum at the Fermi level.

For small values of \(\tAB\) the spectral gap results from the local Hubbard physics.
Disorder then plays a minor role and mostly modifies the bandwidth and therefore the gap size.
An increase of \(\tAB\) leads to a larger bandwidth compared to that of the CPA+DMFT result for \(\tAB=1\).
For larger values of \(\tAB\) the spectral function of the component \(\CmpA\)
is seen to open a pseudogap around the Fermi level
which is accompanied by an increase of spectral weight of the component \(\CmpB\).
For large \(\tAB\), the pseudogap is a result of the off-diagonal disorder.

\subsection{Alloy components with different interaction strengths}
In a binary alloy the strength of the interaction between electrons may also depend on the alloy component.
Therefore we explore the effect of off-diagonal disorder in this case.
We illustrate the results for an extreme case,
namely for a strong repulsion \(\UB=3D\) of the majority component \(\CmpB\) only,
while the minority component remains non-interacting (\(\UA=0\)).
We consider half-filling with \(\vA = 0\), \(\vB = -\UB/2 = -1.5D\) and note that,
in spite of the different values \(\vA \neq \vB\),
the effective (diagonal) disorder strength is 0,
since the Hartree self-energy compensates the difference.
\begin{figure}[htb]
  \includegraphics[width=\linewidth]{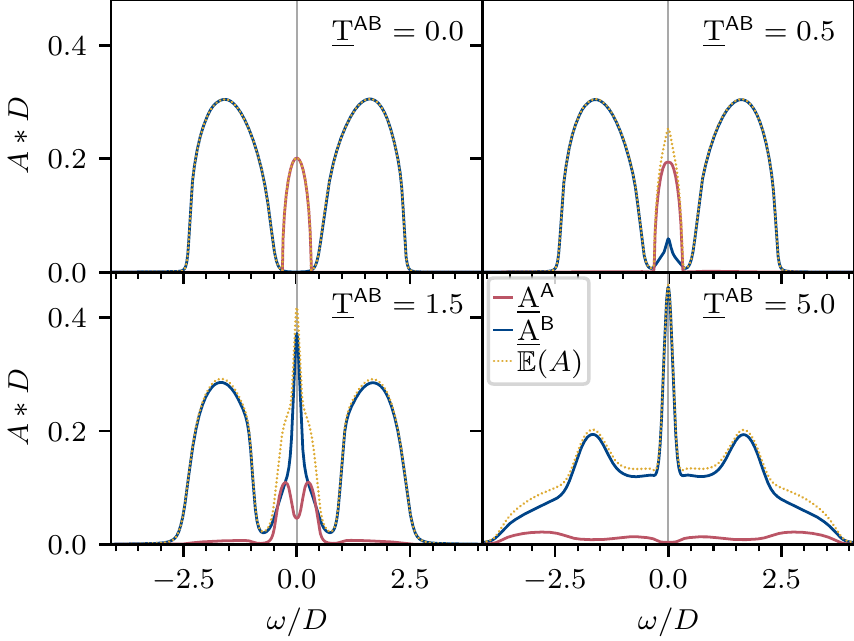}
  \caption{%
    Comparison of spectral functions for different values of the dimensionless hopping parameter \(\tAB\) with \(\UA=0\), \(\UB=3 D\), \(\vA=0\), \(\vB=-1.5D\), \(\cA=0.1=1-\cB\), \(\tAA=\tBB=1\) at \(T=0\).
  }\label{fig:diffU}
\end{figure}
\Cref{fig:diffU} shows the evolution of the spectral function for increasing \(\tAB\).
For \(\tAB=0\), the \(\CmpA\) alloy component is metallic,
while due to the large \(\UB\) value the \(\CmpB\) component is insulating.
We note that the two components have different effective bandwidths due to the different concentrations.
The panel with \(\tAB=0.5\) shows a small peak for the \(\CmpB\) component,
in spite of the large interaction strength.
At \(\tAB=1\), the \(\CmpA\)-\(\CmpA\), \(\CmpA\)-\(\CmpB\), and \(\CmpB\)-\(\CmpB\) hopping probabilities are the same,
which leads to the same effective bandwidths,
and to the appearance of the metallic state for both alloy components.
In \cref{fig:diffU} the panel with \(\tAB=1.5\) shows a distinct peak for the majority component \(\CmpB\) at the Fermi level,
which reduces the spectral function for the \(\CmpA\) component at the Fermi level,
leading to a local minimum.
Increasing the inter-component hopping to \(\tAB=5.0\),
the peak of \(\CmpB\) becomes even larger,
and the spectral weight of \(\CmpA\) almost vanishes at the Fermi level.
The panels with \(\tAB=5.0\) of \cref{fig:sameU,fig:diffU} are seen to be very similar; apparently the interaction of the minority component has little effect on the spectral function.
\begin{figure}[htb]
  \includegraphics[width=\linewidth]{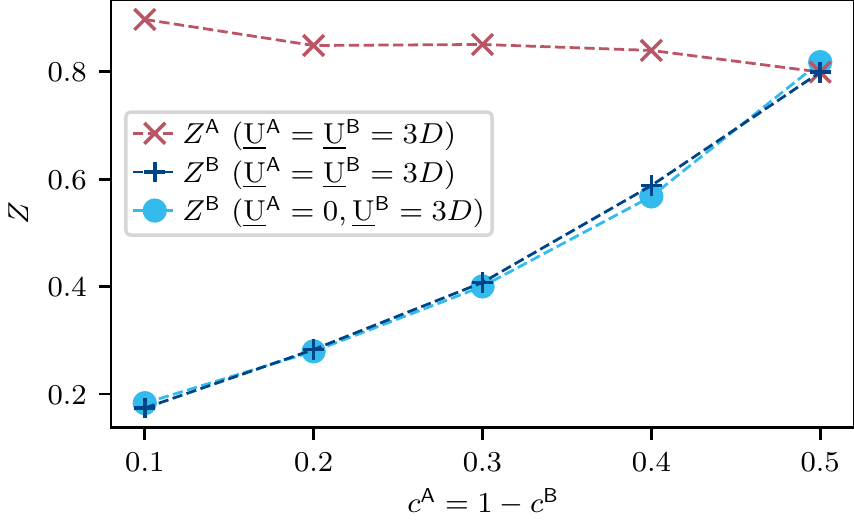}
  \caption{%
    Quasi-particle weight \(Z\) corresponding to \(\tAB=5.0\) in \cref{fig:sameU,fig:diffU} for parameters
    \(\vA=-\UA/2\), \(\vB=-\UB/2\), \(\tAA=\tBB=1\), \(\tAB=5.0\) at  \(T=0\),
    calculated for a shift \(\eta = 0.12\).
  }\label{fig:QPweight}
\end{figure}
\Cref{fig:QPweight} shows the quasi-particle weight
\begin{equation}
  Z^{\alpha} = {\left[1 - \evaluated{\pdv{\Re \Sigma^{\alpha}(\omega + i\eta)}{\omega}}_{\omega=0}\right]}^{-1}
\end{equation}
corresponding to the spectral functions of panel \(\tAB=5.0\) in the \cref{fig:sameU,fig:diffU}.
In spite of the large value of \(\UA=3D\),
the quasiparticle weight \(Z^{\CmpA}\) is large,
with a magnitude around \num{0.9}.
This gives an indication why the panels \(\tAB=5.0\) of \cref{fig:sameU,fig:diffU} are so similar;
the large value of \(\UA=3D\) leads only to a small mass renormalization.
Increasing the concentration of the weakly correlated component \(\CmpA\) leads to a significant increasing of the quasiparticle weight \(Z^{\CmpB}\) for both setups.
This can be explained by the increasing number of \(\CmpA\)-\(\CmpB\) bonds,
which leads to increased mobility of the hopping due to the large value of \(\tAB=5.0\)
compared to the inter-component hoppings \(\tAA=\tBB=1\).

\subsection{Combined effect of diagonal and off-diagonal disorder}
In the following, we explore the combined effect of both diagonal and off-diagonal disorder,
and their interplay with interaction.
We choose a uniform interaction strength \(\UA = \UB = U\)
and introduce diagonal disorder with on-site potentials \(\vA = -1.5D - U/2\), \(\vB = +1.5 D - U/2\).
This means that the scattering strength is of the magnitude \(\delta = (\vB - \vA)/D = 3\).
We consider components with equal bandwidth \(\tAA = \tBB = 1\)
and equal concentration \(c^{\CmpA} = c^{\CmpB} = 0.5\).
Thus, the components are particle-hole conjugate and fulfill the relation
\begin{equation}\label{eq:ph-symmetry}
  \Gloc^{\CmpA\CmpA}(z)
  = -{[\Gloc^{\CmpB\CmpB}(z)]}^{*}.
\end{equation}
\begin{figure*}[thbp]
  \includegraphics[width=\linewidth]{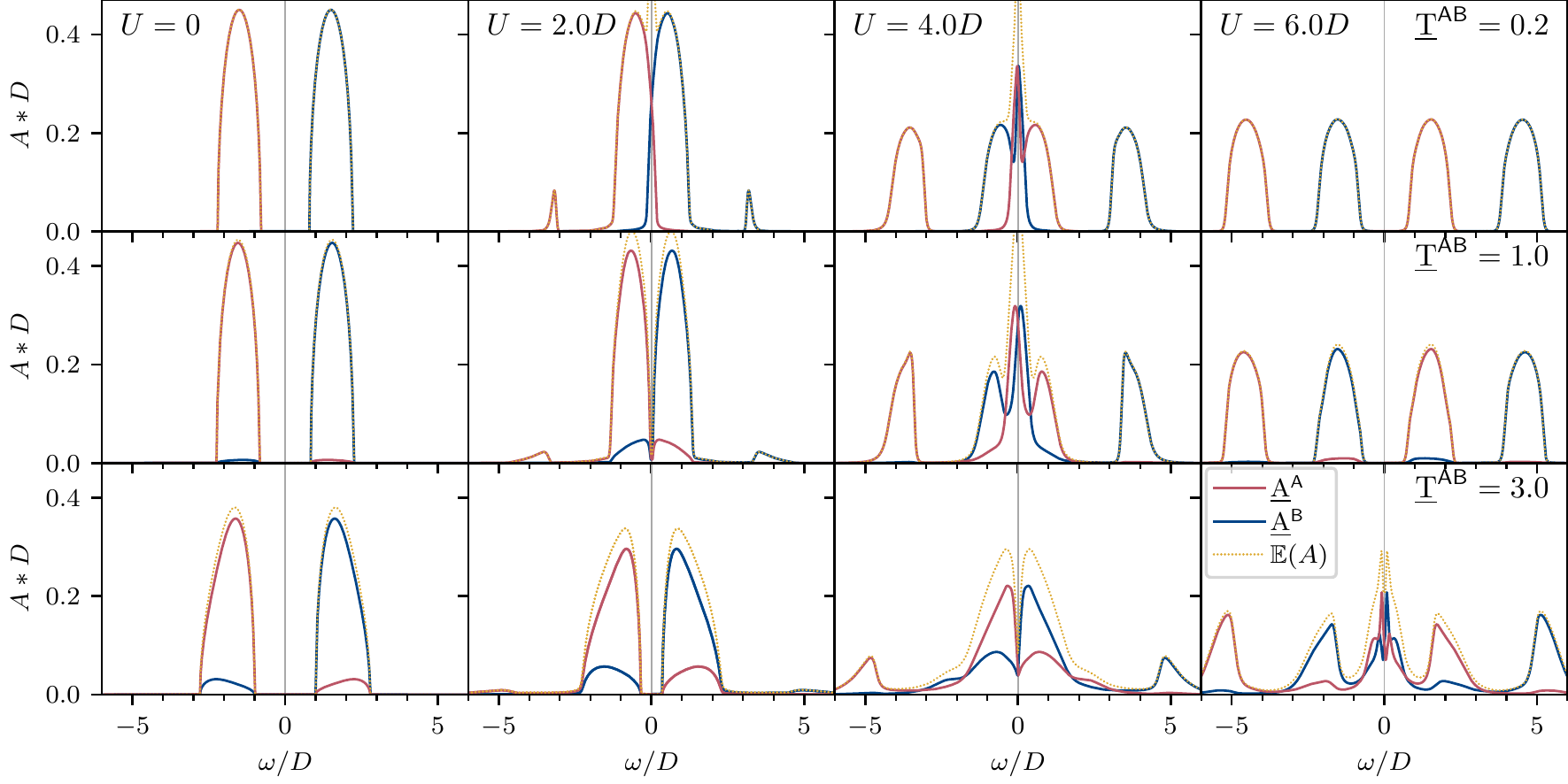}
  \caption{%
    Comparison of the spectral functions for \(\UA=\UB=U\), \(-(\vA + U/2) = \vB + U/2= 1.5 D\), \(c^\CmpA=0.5=c^{\CmpB}\), \(\tAA=\tBB=1\), \(T=0\) for different values of \(\tAB\) and \(U\).
  }\label{fig:ddodd}
\end{figure*}
\Cref{fig:ddodd} shows the average spectral function as well as that for the individual components for different values of the Hubbard parameter \(U\) and dimensionless inter-component hopping amplitudes \(\tAB\).
In the case \(U=0\) (first column of \cref{fig:ddodd}) one starts from the split-band limit.
We first investigate the CPA limit \(\tAB = 1\) (second row of \cref{fig:ddodd}).
In the split band limit, there are no correlation effects:
One component is basically filled \(n_{\sigma}^{\CmpA} \approx 1\), the other is depleted \(n_{\sigma}^{\CmpB} \approx0\).
However, the Hartree energy, \(\Sigma^{\text{H}}_{\sigma} = n_{\sigma} U\), decreases the effective disorder strength
\begin{equation}
\begin{aligned}
  \delta_{\text{eff}}
  &= \frac{(\vB + n_{\sigma}^{\CmpB}U) - (\vA + n^{\CmpA} U)}{D}
  \\%%
  &\approx [\vB - \vA - U]/D
  = \delta - U/D.
\end{aligned}
\end{equation}
Switching on the interaction \(U\) effectively decreases the scattering strength \(\delta\).
From \(U \approx 2 D\) on,
the split-band limit at large scattering strength no longer applies, i.e.,
there is a combination of disorder and interaction effects.
For \(U = 4D\), we see the upper and lower Hubbard bands for each component,
as well as a quasiparticle peak at the Fermi level.
For even larger interaction strength (\(U=6D\))
a Mott insulating phase is observed.
Thus, by increasing \(U\) it is possible to tune the system from an alloy-band insulator, through a metallic phase, to a Mott insulating state.
Similar results were reported by Lombardo \emph{et al.}~\cite{lo.ha.06} for diagonal disorder using CPA+DMFT
for somewhat different parameters and a finite-temperature impurity solver.
The behavior obtained in the CPA limit can now be modified by varying \(\tAB\).
At \(U = 2D\) an off-diagonal hopping \(\tAB < 1\) leads to metallic behavior,
while  \(\tAB > 1\) favors a band gap.
On the other hand, for \(U=6D\) 
a large \(\tAB\) favors metallicity.
For \(\tAB \leq 1\) the spectral function is gapped --- similar to the result obtained in the Hubbard-I approximation~\cite{hubb.63}.

\section{Summary}\label{sec:summary}
We presented a methodological framework for the study of interacting electrons in the presence of diagonal and off-diagonal disorder.
The formalism allows one to explore 
a multi-component system of electrons with random on-site interactions and/or potentials as well as random hopping amplitudes.
For this purpose the Blackman, Esterling, Berk (BEB) formalism for averaging over off-diagonal disorder was combined with the dynamical mean-field theory (DMFT).
We introduced a tensor notation inspired by Koepernik \emph{et al.}~\cite{ko.ve.97,ko.ve.98}, by which the randomness of the components 
is transferred to local indicator tensors defined in an extended space.
In this representation,
the problem can be solved in a single-site approximation, analogous to the coherent potential approximation (CPA).
The computational procedure, including the two-fold self-consistency and the impurity solver\cite{ba.zi.17}, were discussed in detail.
In the limit of zero intercomponent-hopping
an analytic solution for the BEB approximation of the non-interacting Bethe lattice with the general coordination \(Z\) was provided.
We established the exact bounds of the spectral function, and showed that these lie within the maximal bound given by the Gershgorin circle theorem\cite{gers.31}.
The first application of the BEB+DMFT formalism in tensor formulation was the computation of spectral functions.
In particular, we discussed the changes in the spectral function for increasing dimensionless hopping amplitude \(\tAB\) and interaction strengths U.
For alloy components with the same local interaction,
we found a pseudo-gap in the spectral function of the alloy component with lower concentration.
For larger values of \(\tAB\) this  is accompanied by quasiparticles of the dominant alloy component.
The pseudo-gap and the quasiparticle character were inferred by analyzing the respective self-energies.
We found a rather similar behavior in the case of alloy components with different interaction strengths.
For the discussion of both diagonal and off-diagonal disorder we studied equal alloy concentrations and equal alloy component bandwidths for a rather large scattering strength $\delta=3D$.
Increasing $U$ while keeping \(\tAB \leq 1.0\) fixed 
the electronic system was found to undergo two transitions: One from the non-correlated alloy insulator to a metallic state, and then from the metal to a correlated (Mott) insulator.
For larger values of \(\tAB\) the formation of quasiparticles signals the appearance of a metallic state, which eventually disappears again within the Mott insulating phase for large enough values of $U$.
Finally, we note that at the CPA+DMFT level our results are in agreement with previous model-Hamiltonian based publications.
As a matter of fact the BEB formulation is applicable to band structure schemes~\cite{ko.ve.97,ko.ve.98,he.he.17} also for non-orthogonal basis sets.
The present formulation can be naturally extended following the CPA+DMFT~\cite{os.vi.18} methodology with similar computational costs.
Work along this line is in progress.

\begin{acknowledgments}
Financial support by the Deutsche Forschungsgemeinschaft through TRR80 (project F6) Project number 107745057 is gratefully acknowledged.
Y.Z. is supported in part by the NSF China Grant No. 12004383 and the Fundamental Research Funds for the Central Universities.
H.T. gratefully acknowledges support from NSF OAC-1931367 and NSF
DMR-1944974 grants. K.M.T. is partially supported by NSF DMR-1728457 and NSF OAC-1931445 grants.
\end{acknowledgments}

\appendix

\section{An explicit example of the extended Hamiltonian matrix}\label{sec:example}
Here we provide an explicit example of the Hamiltonian matrix \(\mat{H}\) and the extended Hamiltonian \(\cmp{H}\).
We consider a small system --- a chain with \(3\) sites (\(1, 2, 3\)) and \(2\) components \(\CmpA\) and \(\CmpB\).
For this chain the Hamiltonian matrix reads
\begin{equation}
  \mat{H}
  = \begin{pmatrix}
    v_{1} & t_{12} & t_{13}
    \\%%
    t_{21} & v_{2} & t_{23}
    \\%%
    t_{31} & t_{32} & v_{3}
  \end{pmatrix} - \I \mu.
\end{equation}
This is a random matrix since it depends on the configuration.
We consider \(c^\CmpA = 1/3\) and \(c^{\CmpB} = 2/3\);
then one possible configuration is \(\CmpA - \CmpB - \CmpB\).
According to \cref{eq:cmp-parameters},
the Hamiltonian matrix of this configuration takes the values
\begin{equation}
  \mat{H}_{\CmpA\CmpB\CmpB}
  = \begin{pmatrix}
    \cmp*{v}^{\CmpA} & \cmp*{t}^{\CmpA\CmpB}(a) & \cmp*{t}^{\CmpA\CmpB}(2a)
    \\%%
    \cmp*{t}^{\CmpB\CmpA}(a) & \cmp*{v}^{\CmpB} & \cmp*{t}^{\CmpB\CmpB}(a)
    \\%%
    \cmp*{t}^{\CmpB\CmpA}(2a) & \cmp*{t}^{\CmpB \CmpB}(a) & \cmp*{v}^{\CmpB}
  \end{pmatrix} - \I \mu,
\end{equation}
where \(a\) is the distance between neighboring sites.
By contrast, the extended Hamiltonian matrix \(\cmp{H}\)
does not depend on the specific configuration.
We can choose a matrix representation of the tensor \(\cmp{H}\) shown in \cref{eq:extendedH},
by grouping the legs \(i\) and \(\alpha\) on the same side.
This is done explicitly defining the combined index \(n = (i, \alpha)\) [\(m = (j, \beta)\)].
We count \(n = 2i - 1 + n_{\alpha}\) with \(n_{\CmpA} = 0\) and \(n_{\CmpB} = 1\).
Then the extended Hamiltonian \(\cmp*{H}^{\alpha \beta}_{ij} = \cmp*{H}_{nm}\) reads
\begin{multline}
  \cmp{H} + \I \mu=
  \\
  \def\diagblock{\mqty{\dmat[0]{{\cmp*{v}^{\CmpA}}, {\cmp*{v}^{\CmpB}}}}}
  \def\hoppingblock{\mqty{\cmp*{t}^{\CmpA\CmpA}(a) & \cmp*{t}^{\CmpA\CmpB}(a) \\ \cmp*{t}^{\CmpB\CmpA}(a) & \cmp*{t}^{\CmpB\CmpB}(a)}}
  \def\hoppingblockk{\mqty{\cmp*{t}^{\CmpA\CmpA}(2a) & \cmp*{t}^{\CmpA\CmpB}(2a) \\ \cmp*{t}^{\CmpB\CmpA}(2a) & \cmp*{t}^{\CmpB\CmpB}(2a)}}
  \pmqty{%
    \diagblock & \hoppingblock & \hoppingblockk
    \\
    \hoppingblock & \diagblock & \hoppingblock
    \\
    \hoppingblockk & \hoppingblock & \diagblock
  }.
\end{multline}
This \(MN \times MN = 6 \times 6\) matrix contains all \(M^{N}\) possible configurations for the \(N=3\) site problem with \(M=2\) components and is independent of the concentrations \(c^{\alpha}\).
A specific configuration can be selected by applying an appropriate indicator tensor \(\cmp{\eta}\).
For the configuration \(\CmpA - \CmpB - \CmpB\), \(\cmp*{\eta}^{\alpha}_{ij} = \cmp*{\eta}_{nj}\) takes the form
\begin{equation}
  \cmp{\eta}^{\transp}_{\CmpA\CmpB\CmpB}
  = \pmqty{%
    1 & 0 & 0 & 0 & 0 & 0
    \\%%
    0 & 0 & 0 & 1 & 0 & 0
    \\%%
    0 & 0 & 0 & 0 & 0 & 1
  },
\end{equation}
and we obtain the Hamiltonian matrix for this configurations from
\begin{equation}
  \mat{H}_{\CmpA\CmpB\CmpB}
  = \cmp{\eta}^{\transp}_{\CmpA\CmpB\CmpB} \cmp{H} \cmp{\eta}_{\CmpA\CmpB\CmpB}.
\end{equation}
While the matrix representation is suitable for performing calculations,
the tensors notation is often clearer.
The elements diagonal in components \(\alpha\) of the extended Hamiltonian read
\begin{equation}
  \cmp{H}^{\alpha \alpha}
  = \pmqty{%
    \cmp*{v}^{\alpha} & \cmp*{t}^{\alpha\alpha}(a) & \cmp*{t}^{\alpha\alpha}(2a)
    \\%%
    \cmp*{t}^{\alpha \alpha}(a) & \cmp*{v}^{\alpha} & \cmp*{t}^{\alpha \alpha}(a)
    \\%%
    \cmp*{t}^{\alpha\alpha}(2a) & \cmp*{t}^{\alpha \alpha}(a) & \cmp*{v}^{\alpha}
    \\%%
  } - \I \mu,
\end{equation}
the off-diagonal elements \(\alpha \neq \beta\) read
\begin{equation}
  \cmp{H}^{\alpha \beta}
  = \pmqty{%
    0 & \cmp*{t}^{\alpha \beta}(a) & \cmp*{t}^{\alpha\beta}(2a)
    \\%%
    \cmp*{t}^{\alpha \beta}(a) & 0 & \cmp*{t}^{\alpha \beta}(a)
    \\%%
    \cmp*{t}^{\alpha\beta}(2a) & \cmp*{t}^{\alpha \beta}(a) & 0
  },
\end{equation}
and the indicator tensor for the configuration \(\CmpA-\CmpB-\CmpB\) is given by the elements
\begin{equation}
  \cmp{\eta}^{\CmpA}_{\CmpA\CmpB\CmpB}
  = \pmqty{\dmat[0]{1,0,0}}
  \qq{and}%
  \cmp{\eta}^{\CmpB}_{\CmpA\CmpB\CmpB}
  = \pmqty{\dmat[0]{0,1,1}}.
\end{equation}

\section{Efficient evaluation of the local Green's function and its inverse}\label{sec:BEB-Gf_implementation}
To solve the BEB self-consistency equation,
we need to repeatedly evaluate the effective local Green's function
\begin{equation}
  \Gloc(z)
  = \frac{1}{N} \sum_{k} {[\cmp{\xi}(z) - \cmpT \epsilon_{k}]}^{-1},
\end{equation}
with \(\cmp{\xi}(z) = \I z - \cmp{S}(z)\),
or rather its inverse \({(\Gloc)}^{-1}\).
A naive evaluation would be computational costly,
since one needs to invert a matrix for every \(k\)-point and every frequency point.
While this is feasible for small matrices,
it has the potential risk of inaccurate \(k\)-summations (or integrations),
especially for a DOS with singularities as for a one-dimensional or square lattice.
Therefore, we employ an algorithm based on the \emph{compact} singular value decomposition (SVD) of the matrix \(\cmpT\)
\begin{equation}\label{eq:T_SVD}
  \cmpT
  = \mat{U \sigma V}^{\dagger}
  = \mat{U \sigma}^{1/2} \, \mat{\sigma}^{1/2} \mat{V}^{\dagger}
  \eqqcolon \tilde{\mat{U}} \tilde{\mat{V}}^{\dagger},
\end{equation}
which we use to split the matrix;
here we partitioned the singular values symmetrically as \(\tilde{\mat{U}} = \mat{U \sigma}^{1/2}\),
\( \tilde{\mat{V}}^{\dagger} = \sigma^{1/2} \mat{V}^{\dagger}\).
It is important to use the compact SVD as we will explicitly use the inverse \(\mat{\sigma}^{-1}\).
Numerically, the need to truncate small singular values arises.
We note that for the binary alloy
the rank-1 case, where the SVD has to be truncated, is given for a hopping matrix of the type
\begin{equation}
  \begin{pmatrix}
    \tAA & \sqrt{\tAA \tBB}
    \\%%
    \sqrt{\tAA \tBB} & \tBB
  \end{pmatrix}
  \!=\! % reduce space to fit in line
  \begin{pmatrix}
    \sqrt{\tAA}
    \\%%
    \sqrt{\tBB}
  \end{pmatrix}
  \!\!\!% reduce space to fit in line
  \begin{pmatrix}
    \sqrt{\tAA} & \sqrt{\tBB}
  \end{pmatrix}.
\end{equation}
This is the structure of the hopping matrix discussed by Shiba~\cite{shib.71}.
Another prominent rank-1 example is the CPA limit with \(\tAA = \tAB = \tBB = 1\).
The matrix inverse \({[\cmp{\xi}(z) - \cmpT \epsilon_k]}^{-1}\) is calculated using the Woodbury matrix identity~\cite{high.02}.
Furthermore, we calculate the eigendecomposition
\begin{equation}\label{eq:eigendecomposition}
  \tilde{\mat{V}}^{\dagger} \cmp{\xi}^{-1}(z) \tilde{\mat{U}}
  = \mat{P}(z)\mat{d}(z)\mat{P}^{-1}(z),
\end{equation}
where \(\mat{d}(z)\) is the diagonal matrix of eigenvalues.
The \(k\)-dependent Green's function can be expressed as
\begin{equation}
  \cmp{G}(z, k)
  = \cmp{\xi}^{-1}
  - \cmp{\xi}^{-1} \tilde{\mat{U}} \mat{P}{\left[\mat{d}-\frac{\I}{\epsilon_{k}}\right]}^{-1} \mat{P}^{-1} \tilde{\mat{V}}^{\dagger} \cmp{\xi}^{-1},
\end{equation}
where we did not write the \(z\)-dependence explicitly.
We note that only the term in the square bracket depends on \(k\).
Since it contains only diagonal matrices, the matrix inverse only involves the reciprocal matrix elements.
We look at a particular diagonal element with the \(z\)-dependent eigenvalue \(\mat{d}_{ii}(z) = \lambda_{i}(z)\)
\begin{equation}
  -{\left[\lambda_{i}(z) - \frac{1}{\epsilon_{k}}\right]}^{-1}
  = \frac{1}{1/ \lambda_{i}(z) - \epsilon_{k}} \frac{1}{\lambda_{i}^{2}(z)} - \frac{1}{\lambda_{i}(z)}.
\end{equation}
It is straightforward to perform the \(k\)-summation,
as we have the standard form of the lattice Hilbert transform
\begin{equation}
  \frac{1}{N} \sum_{k} \frac{1}{z - \epsilon_{k}}
  \eqqcolon g_{0}(z),
\end{equation}
evaluated at \(1/ \lambda_{i}\).
For simple lattices like the Bethe lattice we know the analytic expression for \(g_{0}\);
a numerical integration can then be avoided.
For the local Green's function we obtain the lengthy expression
\begin{equation}\label{eq:gloc_general}
  \Gloc
  = \cmp{\xi}^{-1}
  + \cmp{\xi}^{-1} \tilde{\mat{U}} \mat{P}{[\mat{d}^{-1} g_{0}(\mat{d}^{-1}) - \I]} \mat{d}^{-1} \mat{P}^{-1} \tilde{\mat{V}}^{\dagger} \cmp{\xi}^{-1},
\end{equation}
where again the \(z\)-dependence was not written out.
For the self-consistency equation of the BEB formalism one only needs the inverse of \(\Gloc\).
The Woodbury matrix identity yields the simpler expression,
\begin{multline}\label{eq:gloc_inv_general}
  \Gloc^{-1}(z) =
  \\%%
  \cmp{\xi}(z)
  + \tilde{\mat{U}} \mat{P}(z) \left[\frac{1}{g_{0}\big(1/\mat{d}(z)\big)} - 1/\mat{d}(z)\right] \mat{P}^{-1}(z) \tilde{\mat{V}}^{\dagger},
\end{multline}
where we explicitly noted the inverse of the diagonal matrices by reciprocal matrix elements.
Considering that the main cost of a naive evaluation of \(\Gloc^{-1}\) arises from the matrix inversion,
this amounts to \(N_{z} (N_{k}+1)\) matrix inversions,
where \(N_{z}\) is the number of frequency points and
\(N_{k}\) is the number of \(k\)-points required for the integration.
The alternative algorithm proposed here requires, on the other hand,
\(N_{z}\) matrix inversions due to the calculation of \(\cmp{\xi}^{-1}(z)\),
and another \(N_{z}\) matrix diagonalizations in the compact space of the SVD\@.
In practice, the calculations were well behaved,
and we encountered no numerical problems regarding the diagonalization of \cref{eq:eigendecomposition}.
For the common case of full rank \(\cmpT\),
we can use the unitarity \(\mat{U}^{\dagger} = \mat{U}^{-1}\) and \(\mat{V}^{\dagger} = \mat{V}^{-1}\)
to simplify the formulas further.
We obtain the simple formulas for the local Green's function
\begin{equation}
  \Gloc(z)
  = \mat{V} \mat{\sigma}^{-1/2} \mat{P}(z) g_{0}\big(1/\mat{d}(z)\big)\mat{P}^{-1}(z) \mat{\sigma}^{-1/2}\mat{U}^{\dagger}
\end{equation}
and its inverse
\begin{equation}
  \Gloc^{-1} (z)
  = \tilde{\mat{U}} \mat{P}(z) \frac{1}{g_{0}\big(1/\mat{d}(z)\big)} \mat{P}^{-1}(z) \tilde{\mat{V}}^{\dagger}.
\end{equation}
Furthermore, we can directly calculate the matrix diagonalization of
\begin{equation}
  \mat{\sigma}^{-1/2} \mat{U}^{\dagger} \cmp{\xi}(z) \mat{V \sigma}^{-1/2}
  = \mat{P}(z) \mat{d}^{-1}(z)\mat{P}^{-1}(z),
\end{equation}
avoiding the need of the \(N_{z}\) matrix inversions for \(\cmp{\xi}(z)\).

\section{BEB self-consistency equation with renormalized indicator tensors}\label{sec:renormalized-BEB}
With \(\alpha \neq \beta\), the high-frequency expansion of the effective medium yields
\begin{align}
  \cmp*{S}^{\alpha \beta}(z)
  &= - \epsilon_{(1)}\cmpT*^{\alpha \beta} + \order{z^{-1} },
  \\%%
  \cmp*{S}^{\alpha \alpha}(z)
  &= \frac{c^{\alpha} - 1}{c^{\alpha}} z
  + \frac{\cmp*{v}^{\alpha} - \mu + \overline{c}^{\alpha}\cmpT*^{\alpha \alpha} \epsilon_{(1)}}{c^{\alpha}}
  + \order{z^{-1}},
\end{align}
where \(\epsilon_{(1)} = \int \mathrm{d} \epsilon \rho(\epsilon) \epsilon\) is the first moment of the DOS,
which vanishes for lattices with a symmetric DOS\@,
\(\cmp*{v}\) incorporates the static part of the self-energy \(\Sigma^{\alpha}(z)\),
and \(\overline{c}^{\alpha} = 1 - c^{\alpha}\) is the concentration complement.
The diagonal of the effective medium, \(\cmp*{S}^{\alpha \alpha}(z)\), has a contribution which grows linearly in \(z\),
and the on-site energies are multiplied by the inverse of the concentration.
The origin of this peculiar structure is evident from \cref{eq:cmpG_physG}
and the definition \cref{eq:BEB_Gloc}.
Unlike the diagonal elements of a one-particle Green's function which behave like \(1/z\) for large \(z\),
the effective local Green's function \(\Gloc(z)\) behaves like \(\mat{c}/z\).
The definition in terms of the effective medium, however, has the regular form of \({[\I z + \dots]}^{-1}\).
This can be resolved by introducing a renormalized version of the component space.
Instead of the indicator tensor \(\cmp{\eta}\), \cref{eq:indicator-tensor},
we use the concentration-scaled indicator tensor
\begin{equation}
  {\cmp*{\gamma}}^{\alpha}_{ij}
  = \sqrt{c^{\alpha}} \cmpind{\alpha}(i) \delta_{ij}
\end{equation}
and the Moore--Penrose inverse~\cite{moor.20,penr.55} \(\cmp{\gamma}^{+}\) (which is in this case the left-inverse, i.e., \(\cmp{\gamma}^{+}\cmp{\gamma} = \I\)) of its equivalent matrix representation.
The components of the Moore--Penrose inverse read
\begin{equation}
  {\cmp*{\gamma}^{+}}^{~\alpha}_{ij}
  = \begin{cases}
    \frac{1}{\sqrt{c^{\alpha}}} \cmpind{\alpha}(i) \delta_{ij} &\qif c^{\alpha} > 0,
    \\%%
    0 &\qif c^{\alpha} = 0.
  \end{cases}
\end{equation}
We can express the projector \cref{eq:def:projector} with the \(\cmp{\gamma}\) tensor:
\begin{equation}
  \cmp{\chi}
  = \cmp{\gamma \gamma}^{+}.
\end{equation}
For the renormalized BEB formalism,
we define the component Green's function and the Hamiltonian matrix in terms of \(\cmp{\gamma}\) and the inverse \(\cmp{\gamma}^{+}\) as
\begin{equation}
\begin{aligned}
  \scmp{G}(z)
  &\coloneqq {(\cmp{\gamma}^{+})}^{\transp} \mat{G}(z) \cmp{\gamma}^{+},
  \\%%
  \mat{H}
  &\eqqcolon {\cmp{\gamma}} \scmp{H} {\cmp{\gamma}}^{\transp}.
\end{aligned}
\end{equation}
Compared to the definitions in \cref{sec:model},
the Green's function and the Hamiltonian are scaled by the concentration.
This can be conveniently demonstrated in the locator expansion
\begin{equation}
  \mat{G}(z)
  = \mat{g}(z) + \mat{g}(z) \mat{T} \mat{G}(z),
\end{equation}
where \(\mat{g}(z) = {[\I z - \mat{v}]}^{-1}\) is the locator and \({(\mat{T})}_{ij} = t_{ij}\).
Sandwiching this equation by \({(\cmp{\gamma}^{+})}^{\transp}\) and \(\cmp{\gamma}^{+}\),
we obtain
\begin{multline}
  {(\cmp{\gamma}^{+})}^{\transp} \mat{G}(z) \cmp{\gamma}^{+}
  ={(\cmp{\gamma}^{+})}^{\transp} \mat{g}(z) \cmp{\gamma}^{+}
  \\
  + {(\cmp{\gamma}^{+})}^{\transp} \mat{g}(z) \cmp{\gamma}^{+}
  {\cmp{\gamma}}\mat{T} \cmp{\gamma}^{\transp}
  {(\cmp{\gamma}^{+})}^{\transp} \mat{G}(z) \cmp{\gamma}^{+},
\end{multline}
where we inserted the identity \(\cmp{\gamma}^{+} \cmp{\gamma} = \I\).
This can be written in terms of the renormalized component quantities
\begin{equation}
  \scmp{G}(z)
  = \scmp{g}(z) + \scmp{g}(z) \scmp{T}\, \scmp{G}(z).
\end{equation}
Compared to the regular BEB formalism
the Green's functions are scaled with the reciprocal concentration,
and the hopping matrix with the concentration \(\scmp*{T}^{\alpha \beta}_{ij} = \sqrt{c^{\alpha}} \cmp*{t}^{\alpha \beta}(\abs{\vb{r}_{i} - \vb{r}_{j}}) \sqrt{c^{\beta}}\).
The renormalized component Green's function relates now to the one-particle Green's function in the following way:
\begin{equation}
\Expval(\scmp*{G}^{\alpha \beta}_{ij})
  = \begin{cases}{}
    \Expval(G_{ii}(z) \vert i \mapsto \alpha) \delta^{\alpha \beta}
    &\qif i = j,
    \\%%
    \sqrt{c^{\alpha} c^{\beta}} \Expval(G_{ii}(z) \vert i,j \mapsto \alpha, \beta).
    &\qif i \neq j
  \end{cases}
\end{equation}
There is no more concentration prefactor for the local Green's function,
which is the central quantity of the BEB formalism, since it is a local theory.
The renormalized version of the self-consistency equation \cref{eq:BEB_self-consistency,eq:ave}
reads
\begin{equation}
  0
  = \scmp{g}_{\text{loc}}^{-1}(z) - \overline{\scmp{g}}^{-1}(z),
\end{equation}
with the diagonal matrix
\begin{equation}
  \overline{\scmp*{g}}^{\alpha \beta}(z)
  = \frac{c^{\alpha} \delta^{\alpha \beta}}
  {{(\scmp{g}_{\text{loc}}^{-1})}^{\alpha \alpha} + \scmp*{S}^{\alpha \alpha} - \overline{c}^{\alpha} z + c^{\alpha}(\mu - \cmp*{v}^{\alpha} - \Sigma^{\alpha})}.
\end{equation}
With \(\alpha \neq \beta\), the high-frequency expansion of the renormalized effective medium yields
\begin{align}
  \scmp*{S}^{\alpha \beta}(z)
  &= - \epsilon_{(1)} \sqrt{c^{\alpha}} \cmpT*^{\alpha \beta} \sqrt{c^{\beta}} + \order{z^{-1}}
  \\%%
  \scmp*{S}^{\alpha \alpha}(z)
  &= \cmp*{v}^{\alpha} - \mu + \overline{c}^{\alpha}\cmpT*^{\alpha \alpha} \epsilon_{(1)} + \order{z^{-1}}.
\end{align}
The scaling removes the contribution proportional to \(z\),
and for a symmetric DOS the static part is simply the on-site energy of the components.
Furthermore, the static part remains finite with vanishing concentration.

\section{Hybridization of the Bethe lattice}\label{sec:Bethe_hybridization}
Specific to the BEB+DMFT scheme is an alloy-component dependent hybridization function \(\Delta^{\alpha}(z)\) computed according to \cref{eq:hyb_BEB-DMFT}.
The hybridization function describes the hopping process into and out-of the impurity site.
The special form of the lattice Hilbert transform \(g_{0}(z)\) for the Bethe lattice
\begin{equation}\label{eq:Bethe-relation}
  z - 1/g_{0}(z) = {({D}/{2})}^{2} g_{0}(z)
\end{equation}
gives a direct relation between the hybridization function \(\Delta^{\alpha}(z)\) and the effective local Green's function \(\Gloc(z)\) \cref{eq:BEB_Gloc}.
We promote the hybridization function \cref{eq:hyb_BEB-DMFT} to a full hybridization matrix
\begin{equation}
  \cmp{\Delta}(z)
  = \I z - \cmp{S}(z) - \Gloc^{-1}(z)
  \eqqcolon \cmp{\xi}(z) - \Gloc^{-1}(z),
\end{equation}
whose diagonal elements are the physical hybridization function \(\cmp*{\Delta}^{\alpha \alpha} = \Delta^{\alpha}\).
Using the representation \cref{eq:gloc_inv_general} given in \cref{sec:BEB-Gf_implementation},
we can apply the identity \cref{eq:Bethe-relation}:
\begin{equation}\label{eq:hybridization-expansion}
  \cmp{\Delta}(z)
  = {(D/2)}^{2} \tilde{\mat{U}} \mat{P}(z) g_{0}\big(1/\mat{d}(z)\big) \mat{P}^{-1}(z) \tilde{\mat{V}}^{\dagger}.
\end{equation}
Likewise, the matrix \(\cmpT\Gloc(z)\cmpT\) can be expressed using \cref{eq:T_SVD,eq:eigendecomposition,eq:gloc_general}:
\begin{equation}\label{eq:TgT-expansion}
  \cmpT \Gloc(z) \cmpT
  = \tilde{\mat{U}} \mat{P}(z) g_{0}\big(1/\mat{d}(z)\big) \mat{P}^{-1}(z) \tilde{\mat{V}}^{\dagger}.
\end{equation}
Thus, comparing \cref{eq:hybridization-expansion,eq:TgT-expansion},
we identify the relation for the Bethe lattice:
\begin{equation}
  \cmp{\Delta}(z)
  = {({D}/{2})}^{2} \cmpT \Gloc(z) \cmpT.
\end{equation}
The diagonal elements are the hybridization function,
which reads
\begin{equation}
  \Delta^{\alpha}(z)
  = {(D/2)}^{2} \sum_{\beta} \abs{\cmpT*^{\alpha \beta}}^{2}\Gloc^{\beta \beta}(z).
\end{equation}
In the CPA case \(\cmpT*^{\alpha \beta} = 1\)
we evidently recover the component independent hybridization:
\begin{equation}
  \Delta^{\alpha}(z)
  = {(D/2)}^{2} \sum_{\alpha} \Gloc^{\alpha \alpha}(z)
  = {(D/2)}^{2} \Expval\big(G(z)\big).
\end{equation}
For the case of a binary alloy shown by \cref{fig:ddodd},
we see that for \(\tAB < 1 = \cmpT*^{\alpha\alpha}\) the hybridization for \(\CmpA\) stems mostly from \(\CmpA\) sites,
while for \(\tAB > 1 = \cmpT*^{\alpha\alpha}\) the main contribution to the hybridization of \(\CmpA\) comes from the \(\CmpB\) sites.

\bibliography{reference}
\end{document}